%% file: main.tex
\documentclass[a4paper,11pt]{article}
\pdfoutput=1 

\usepackage{jinstpub} 
\usepackage{wasysym}
\usepackage[separate-uncertainty=true,product-units=power,per-mode=symbol]{siunitx}
\usepackage{circuitikz}
\usepackage{todonotes}
\usepackage{algorithm2e}
\usepackage{xcolor}

\definecolor{MPLBlue}{HTML}{1f77b4}
\definecolor{MPLOrange}{HTML}{ff7f0e}
\definecolor{MPLGreen}{HTML}{2ca02c}

\graphicspath{{img/}}

\title{Extending the dynamic range of SiPMs by understanding their non-linear behavior}

\author{T.~Bretz,}
\author{T.~Hebbeker,}
\author[1]{and J.~Kemp\note{Corresponding author.}}
\affiliation{III. Physikalisches Institut A, RWTH Aachen University \\ Otto-Blumenthal-Stra\ss{}e, 52074 Aachen, Germany}

\emailAdd{kemp@physik.rwth-aachen.de}

\abstract{This publication focuses on the study of silicon photomultipliers (SiPMs) in view of a reconstruction of the incident photon flux in the regime of highly non-linear response. SiPMs are semiconductor based light detectors compiled of avalanche photodiodes operated in Geiger mode. They are both mechanically and optically very robust and have a high gain and photon detection efficiency. These features make them ideal photonsensors in a wide range of applications and they are nowadays replacing conventional photomultiplier tubes in many experiments. The cellular structure of SiPMs where each cell can only detect one photon at a time results in a non-linear dynamic range limiting the possible applications. 

We studied a commonly used SiPM model based on an equivalent electronic circuit that allows the simulation of the SiPM response in many situations. Dedicated measurements with two consecutive light pulses prove its applicability. By adapting the model to the measurements, intrinsic parameters of the SiPM such as quenching resistance or diode capacitance can be determined. With the obtained intrinsic parameters, the model correctly describes the recharge behavior of the SiPM cells.

Based on the model, an algorithm was developed to correct the non-linearity of the dynamic range of SiPMs. As the model contains full information on the recharge of the SiPM cells, the effects leading to the non-linearity can be corrected for. The algorithm exploits the time information in the measured voltage signal and reconstructs the number of incident photons. It has shown an excellent performance and allows to increase the dynamic range with only \SI{10}{\percent} deviation from linearity by at least two orders of magnitude.
}


\newcommand{\Vbi}{V_\mathrm{bi}}
\newcommand{\Vbd}{V_\mathrm{bd}}
\newcommand{\Vov}{V_\mathrm{ov}}
\newcommand{\Rq}{R_\mathrm{q}}
\newcommand{\Cd}{C_\mathrm{d}}
\newcommand{\Rs}{R_\mathrm{s}}
\newcommand{\Cq}{C_\mathrm{q}}
\newcommand{\Id}{I_\mathrm{d}}
\newcommand{\Cg}{C_\mathrm{g}}
\newcommand{\Vo}{V_\mathrm{meas}}
\newcommand{\Vt}{V_\mathrm{t}}
\newcommand{\Vut}{V_\mathrm{ut}}
\newcommand{\Vovinst}{\tilde{V}_\mathrm{ov}}
\newcommand{\Vspe}{V_\mathrm{spe}}
\newcommand{\pxt}{p_\mathrm{xt}}
\newcommand{\Sdel}{S_\mathrm{del}}
\newcommand{\Nmeas}{N_\mathrm{meas}}

\newcommand{\Ncell}{N_\mathrm{cell}}
\newcommand{\NgammaOft}{N_\gamma(t)}
\newcommand{\NgammaOftau}{N_\gamma(\tau)}
\newcommand{\NgammaPrimeOft}{N'_\gamma(t)}
\newcommand{\NgammaPrimeOftau}{N'_\gamma(\tau)}
\newcommand{\SOft}{S(t)}
\newcommand{\SPEOft}{\textit{SPE}(t)}
\newcommand{\SPERealOft}{\textit{SPE}_\mathrm{real}(t)}
\newcommand{\noiseOft}{n(t)}
\newcommand{\SPEOftAndTau}{\textit{SPE}(t|\tau)}
\newcommand{\QE}{\textit{QE}}
\newcommand{\PT}{P_\mathrm{T}}
\newcommand{\FF}{\textit{FF}}
\newcommand{\Vone}{V_1}
\newcommand{\Vtwo}{V_2}
\newcommand{\Vthree}{V_3}
\newcommand{\pemit}{p_\mathrm{emit}}

\begin{document}
\maketitle
\flushbottom

\section{Introduction}
Silicon photomultipliers (SiPMs) are well understood photon detectors and considered in more and more applications as a replacement of conventional photomultiplier tubes (PMTs)~\cite{SSDSiPMModule,FACT,IceTopUpgrade}. Their advantages range from tremendous optical properties such as high photon detection efficiency and robustness with respect to high light fluxes to mechanical robustness and insensitivity to magnetic fields. SiPMs are being manufactured with great precision leading to an outstanding photon counting resolution and only small variations between different devices of the same type.

Many successful attempts and techniques were developed to simulate the response of SiPMs or to measure intrinsic parameters. In this publication, we will discuss a novel approach based on a simple simulation that exploits an equivalent electronic circuit of SiPMs. It allows to measure intrinsic parameters in a dedicated setup and to make predictions for other measurements such as the dynamic range.

The response of SiPMs is intrinsically non-linear with respect to the number of impinging photons due to their cellular structure and depends on the temporal distribution of the incident photons. In most applications the light flux is low and a linear approximation is used. We will show that a reconstruction of the incident number of photons is possible even in the regime where the response deviates from linearity by one order of magnitude. The developed algorithm exploits the time dependency of the measured voltage signal of the SiPM and makes use of the aforementioned electronic model. It allows to use SiPMs also in applications where a precise measurement of the incident number of photons is necessary over a wide dynamic range.

\section{Silicon photomultipliers}
Silicon photomultipliers (SiPMs) are semiconductor based light sensitive devices with a high gain that allows the detection of single photons. In recent years, their development has made significant progress leading to a performance exceeding that of conventional photomultiplier tubes (PMTs) in many aspects. Several publications give a good overview of the working principle and characteristics of SiPMs~\cite{AcerbiSiPMOverview,Klanner:2018ydn}. Here, only a brief introduction of the details necessary for the understanding of the following work will be given.

SiPMs are composed of avalanche photodiodes operated in Geiger mode (G-APDs). Each G-APD is connected in series to a quenching resistor and this combination is referred to as \emph{cell}. Typical cell pitches range from \SIrange{10}{100}{\micro\meter} on SiPMs of size \SIrange{1}{36}{\square\milli\meter}~\cite{HamamatsuS13360,SensLCSeries}. They are operated with a reverse bias voltage $\Vbi$ which is bigger than the intrinsic breakdown voltage $\Vbd$ of the G-APD. The overvoltage $\Vov=\Vbi-\Vbd$ is a characteristic parameter that properties of the SiPM such as photon detection efficiency (PDE), gain or crosstalk probability depend on. The PDE is given by the product of quantum efficiency $\QE$, the avalanche triggering probability $\PT$ and the geometrical fill factor $\FF$:
\begin{equation}
    \textit{PDE}(\Vov) = \QE\cdot\PT(\Vov)\cdot\FF
\end{equation}
where only the avalanche triggering probability depends on the applied overvoltage, $\QE$ is the probability for a photon to reach the region of the silicon where triggering is possible and $\FF$ is the fraction of sensitive area compared to the geometrical area of the SiPM.

The breakdown voltage depends on the specific device as well as on the type of the SiPM. It is typically well below \SI{100}{\volt} with $\Vov<\SI{10}{\volt}$.

When a photon impinges on an SiPM cell, an avalanche of charge carriers can be created leading to a discharge of the cell. The avalanche is quenched by the quenching resistor connected in series. The amount of released charge and the shape of the electric pulse are characteristic for the SiPM and referred to as photoelectron equivalent (p.e.). When the charge of a p.e.~is known, the response of the SiPM can be expressed in terms of p.e.~which, in case of simultaneously impinging photons, corresponds to the number of cells that broke down. After a discharge, the cell needs to recharge again with a characteristic time constant. This time constant depends on the intrinsic parameters of the SiPM such as diode capacitance and quenching resistor but also on the readout impedance and will be introduced in more detail in section~\ref{sec:SiPMElectricModel}. During this time period, the applied overvoltage is reduced affecting also the gain and avalanche triggering probability of the cell. The cellular structure as well as the binary nature of the photon detection of a single cell lead to an intrinsically non-linear response. This effect will be discussed in detail in section~\ref{sec:DynamicRangeExtension} and an algorithm will be presented that allows to recover the number of incident photons despite the non-linear response.

On the other hand, the cellular structure and the simple working principle of a single cell allow the implementation of realistic simulations of their response. Comparing the simulations to measurements allows for extraction of intrinsic parameters of SiPMs such as the capacitance of the diode of the cell $\Cd$ or the quenching resistance $\Rq$. Such a simulation and measurement will be introduced in sections~\ref{sec:Simulation} and \ref{sec:IntrinsicParameterMeasurement}, respectively.

The different noise phenomena of SiPMs are often separated into \emph{correlated} and \emph{thermal} noise. Thermal noise originates from thermal excitation of electrons in the silicon lattice. This process is indistinguishable from the excitation initiated by an incident photon. The rate of breakdowns due to thermal excitation scales exponentially with the temperature and doubles roughly every \SI{8}{\kelvin}~\cite{RenkerSiPMOverview}. Correlated noise is split into crosstalk and afterpulsing. For crosstalk, recombinations of electron-hole pairs in the avalanche result in the emission of photons. These photons can traverse the SiPM and initiate breakdowns at neighboring cells. Afterpulsing originates from impurities in the silicon which trap electrons or holes of the avalanche. They get released with a delay and initiate a second avalanche. Afterpulsing and crosstalk are therefore always connected to a breakdown of a cell which can be due to thermal noise or due to an impinging photon. In recent devices, crosstalk is limited to a few percent and afterpulsing only occurs at the level of a few per mill~\cite{HamamatsuS13360,SensLCSeries}.

\section{Model description}\label{sec:Simulation}
We aim at designing a simulation in particular for the application when the SiPM is exposed to a bright flash of light and its response is highly non-linear. This means that the simulation of the effects resulting from a single photon should be fast. Thermal noise is not considered as it can be neglected in case of a high light flux. The focus is placed on the recharge behavior of the cells which must be modeled precisely.

Simulations of SiPMs were performed by many groups with focus on different aspects of their characteristics. In many cases, the impinging photon is tracked through the silicon and all optical and electrical effects are taken into account~\cite{G4SiPM,AcerbiSiPMOverview,FACTCalibration}. Here, a different approach is used based on an equivalent electronic model of SiPMs. It allows to simulate the electric pulse measured for a given temporal and spatial distribution of impinging photons. The photons are not tracked through the silicon but only the produced electric pulse is determined from a few analytic equations that will be introduced in the following section. This kind of simulation is faster than those tracking the photons because all the physical interactions in the silicon are not simulated.

\subsection{SiPM equivalent electronic model}\label{sec:SiPMElectricModel}
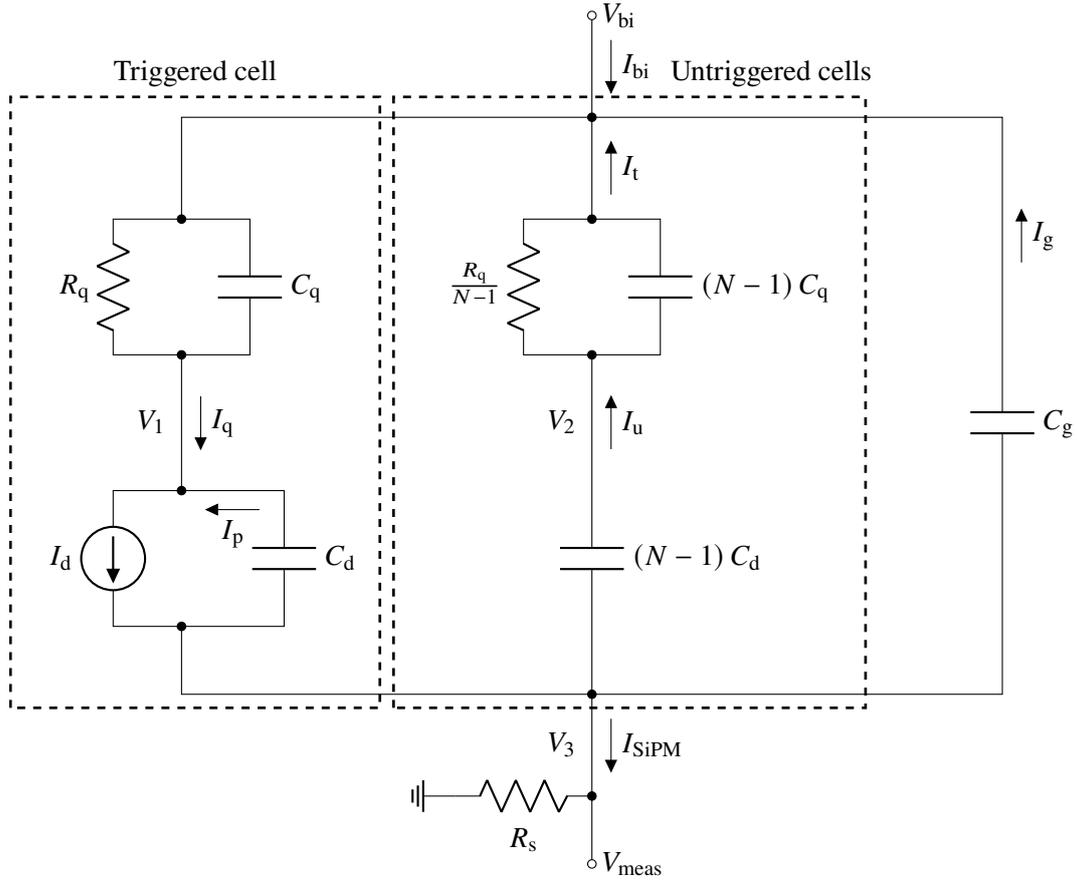
\begin{figure}
    \centering
    \input{img/SiPMCircuitDiagram.tex}
    \caption{The equivalent electronic circuit of an SiPM according to~\cite{CorsiSiPMSignalSource} when being read out over a shunt resistor $\Rs$. The SiPM with $N$ cells is split in two main contributions. One triggered cell is represented by the quenching resistor $\Rq$ and its capacitance $\Cq$ connected in series to a current source $\Id$ at the diode with capacitance $\Cd$. The remaining $N-1$ untriggered cells are only represented by a single cell. Its quenching resistance, quenching capacitance and diode capacitance are modified according to the number of cells connected in parallel and no current source is present. An additional parasitic grid capacitance $\Cg$ is connected in parallel to the triggered and untriggered cell. Taken from~\cite{ThesisJulian}. Values for the different parameters are given in table~\ref{tab:6025PEIntrinsicParameters} for an SiPM of type Hamamatsu S13360-6025PE~\cite{HamamatsuS13360}.}
    \label{fig:SiPMElectricCircuit}
\end{figure}

An electric model of G-APDs was proposed in~\cite{Cova} and later extended to full SiPMs which are multiple G-APDs connected in parallel by~\cite{CorsiSiPMElectricalCharacterization,CorsiSiPMSignalSource}. An equivalent electronic circuit of SiPMs based on these works is shown in figure~\ref{fig:SiPMElectricCircuit}.
\begin{table}\centering
\begin{tabular}{|c|c|c|c|c|c|}\hline
	$N_\mathrm{cells}$ & $R_\mathrm{q}$ / \SI{}{\kilo\ohm} & $C_\mathrm{d}$ / fF & $C_\mathrm{q}$ / fF & $C_\mathrm{g}$ / pF & gain\\\hline
    57600 & 750 & 20.6 & 1.6 & 41.1 & \num{7e5}\\\hline
\end{tabular}
	\caption{Values of the capacitances and resistors according to the schematics given in figure~\ref{fig:SiPMElectricCircuit} for an SiPM of type Hamamatsu S13360-6025PE~\cite{HamamatsuS13360,HamamatsuPrivateCommunication}. The variations between different sensors of the same type are \SI{10}{\percent} for the capacitances and \SI{20}{\percent} for the quenching resistor~\cite{HamamatsuPrivateCommunication}. These uncertainties correspond to the maximum possible and not to Gaussian standard deviations.}
	\label{tab:6025PEIntrinsicParameters}
\end{table}
Throughout this publication an SiPM of type Hamamatsu S13360-6025PE~\cite{HamamatsuS13360} is considered. Its intrinsic parameters are given in table~\ref{tab:6025PEIntrinsicParameters}. The SiPM is biased at a voltage $\Vbi$ and readout over a resistor $\Rs$ where a voltage $\Vo$ is measured. It should be noted that the intrinsic breakdown voltage of the cells corresponds to a constant voltage source that is connected in series in the cells. It can thus be neglected here so that $\Vbi$ is equal to the overvoltage $\Vov$.

To understand the behavior of the SiPM after the triggering of a single cell, the voltages $\Vone$, $\Vtwo$ and $\Vthree=\Vo$ are of interest. The voltage $\Vthree$ is measured at the readout system. The voltage $\Vone-\Vthree$ drops over the diode of the triggered cell and $\Vtwo-\Vthree$ over the diode of the untriggered cell. The voltages $\Vone$ and $\Vtwo$ have an impact on the behavior of the cell for a consecutive second impinging photon. The response of the cells for a consecutive second breakdown thus depends on the time evolution of these three voltages. The time evolution will be calculated in the following and allows a full electrical simulation of the SiPM.

Generally, the gain $g$ and released charge $Q$ for a triggered avalanche are approximated~\cite{AcerbiSiPMOverview} as
\begin{equation}\label{eq:SiPMGain}
    g = \frac{Q}{e} = \frac{\Vov\cdot\left(\Cd+\Cq\right)}{e}.
\end{equation}
The resistance of the diode of the cell is negligible compared to the quenching resistor resulting in a very fast discharge of the cell of $\mathcal{O}\left(\mathrm{ps}\right)$ compared to $\mathcal{O}\left(\mathrm{ns}\right)$ for the recharge through the quenching resistor. Hence, a valid simplification can be made by assuming the diode current as being an infinitely short pulse:
\begin{equation}
    \Id(t) = Q\cdot\delta(t).
\end{equation}

The time evolution of the voltages $\Vone$, $\Vtwo$ and $\Vthree$ has been studied in detail in several publications~\cite{MaranoSiPMAnalyticalAnalysis,Jha:2013,MasterThesisJohannes,ThesisJulian} and will therefore not be introduced in full detail here. The differential equations for the currents across the circuit are given in appendix~\ref{app:DifferentialEquations}. Details on how to solve them can be found in the aforementioned publications.

\begin{figure}
    \centering
    \includegraphics[width=\textwidth]{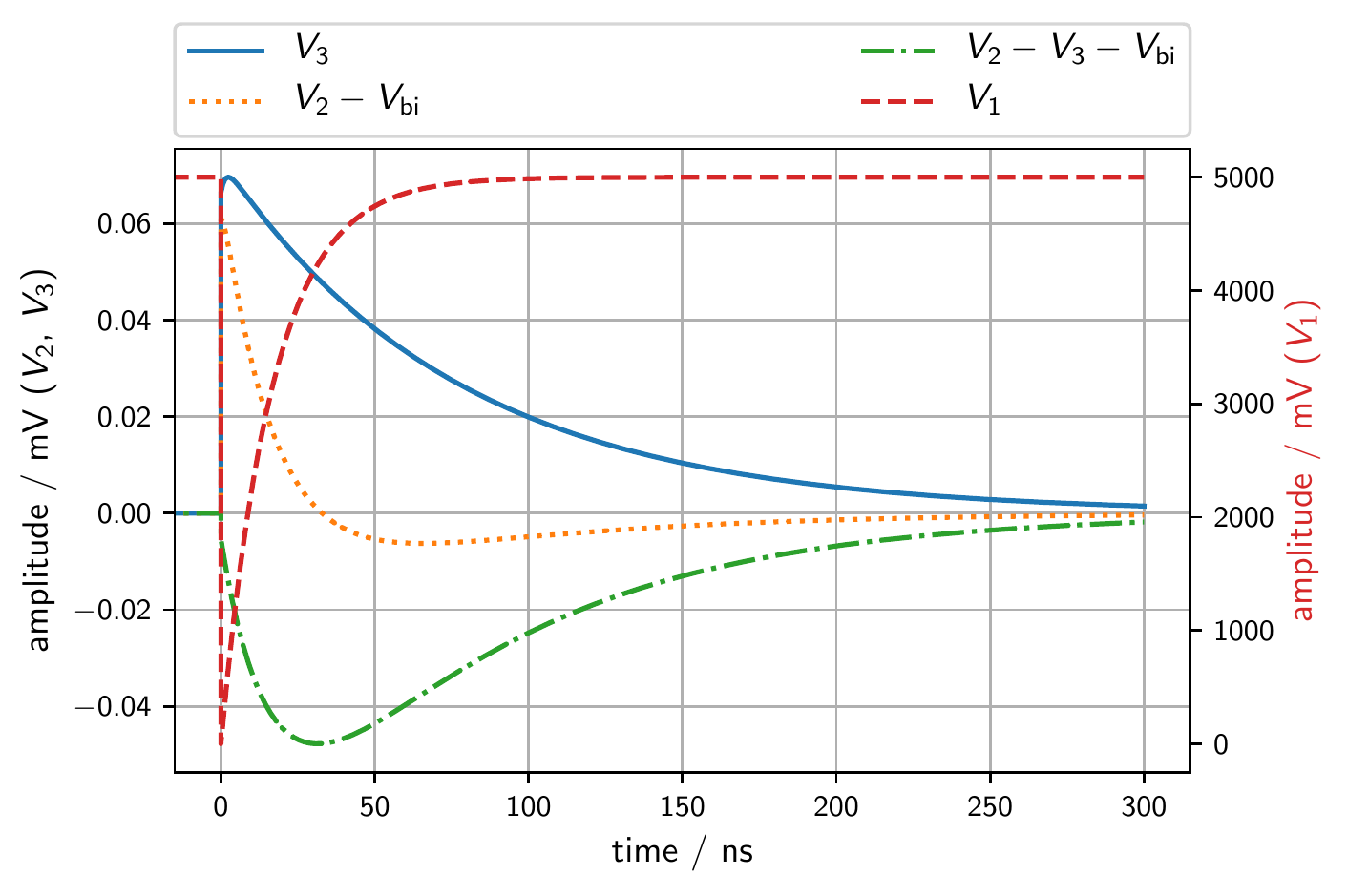}
    \caption{Time evolution of the different voltages at the SiPM as calculated for the electronic circuit shown in figure~\ref{fig:SiPMElectricCircuit} for an applied overvoltage of $\Vov=\Vbi=\SI{5}{\volt}$. An impedance of $\Rs=\SI{50}{\ohm}$ was used. The intrinsic parameters are given in table~\ref{tab:6025PEIntrinsicParameters} and correspond to an SiPM of type Hamamatsu S13360-6025PE. Note the different y-scales. Taken from~\cite{ThesisJulian}.}
    \label{fig:DoublePulseVoltageRecovery}
\end{figure}

The following results for the time evolution of the voltages across the SiPM are obtained:
\begin{align}\label{eq:SiPMVoltageRecovery}
\begin{split}
	\Vone(t) &= \Vbi\underbrace{-\frac{Q\Rq}{N}\left(\frac{N-1}{\tau_\text{q}}\text{e}^{-t/\tau_\text{q}}+\frac{\Rs\Cg}{c_2}\left(A_1\text{e}^{-t/\tau_-}+\left(1-A_1\right)\text{e}^{-t/\tau_+}\right)\right)}_{\tilde{\Vone}(t)}\\
	\Vtwo(t) &= \Vone+\frac{Q\Rq}{\tau_\text{q}}\text{e}^{-t/\tau_\text{q}}\\ &= \Vbi\underbrace{-\frac{Q\Rq}{N}\left(-\frac{1}{\tau_\text{q}}\text{e}^{-t/\tau_\text{q}}+\frac{\Rs\Cg}{c_2}\left(A_1\text{e}^{-t/\tau_-}+\left(1-A_1\right)\text{e}^{-t/\tau_+}\right)\right)}_{\tilde{\Vtwo}(t)}\\
	\Vthree(t) &= \frac{Q\Rs\Rq\Cq}{c_2}\left(A_2\text{e}^{-t/\tau_-}+\left(1-A_2\right)\text{e}^{-t/\tau_+}\right)
\end{split}
\end{align}
with
\begin{align}\label{eq:SiPMTimeConstants}
\begin{split}
	\tau_\text{q} &= \Rq\left(\Cd+\Cq\right)\\
	\tau_\pm &= \frac{2c_2}{c_1\pm\sqrt{c_1^2-4c_2}}\\
	A_1 &= \frac{1}{2}+\frac{2c_2/\left(\Rs\Cg\right)-c_1}{2\sqrt{c_1^2-4c_2}}\\
	A_2 &= \frac{1}{2}+\frac{2c_2/\left(\Rq\Cq\right)-c_1}{2\sqrt{c_1^2-4c_2}}
\end{split}
\end{align}
and
\begin{align}\label{eq:SiPMRecoveryConstants}
\begin{split}
	c_1 &= \Cd \left( N \Rs + \Rq \right) + \Cg \Rs + \Cq \Rq \\
	c_2 &= \Rs\Rq\left(\Cg\Cq+\Cd\left(\Cg+N \Cq\right)\right).
\end{split}
\end{align}
A total of three different time constants for the exponential recharge of the cells can be identified, $\tau_\mathrm{q}$, $\tau_+$ and $\tau_-$. The time constant $\tau_\mathrm{q}$ originates from the current from the bias supply through the quenching resistor. The time constants $\tau_\pm$ describe the measured signal $\Vthree$. Depending on the exact values of the intrinsic resistors and capacitors, they describe a fast rising or falling component and a slow falling component. For the SiPM from the example, $A_2>1$ is true and $V_3(t)$ has a fast rising component with time constant $\tau_+$ and a slow falling component with $\tau_-$. The three voltages are depicted in figure~\ref{fig:DoublePulseVoltageRecovery} for the SiPM from the example introduced earlier.

The time evolution of $\Vone$, $\Vtwo$ and $\Vthree$ shows a step like behavior around $t=0$ when the breakdown occurs. This is a result of the assumption of an infinitely short current $\Id$.

It should be noted that the voltage drop over the diodes of the cells is $\Vt=\Vone-\Vthree$ and $\Vut=\Vtwo-\Vthree$ for the triggered and untriggered cells, respectively. After the breakdown, these voltages need to recover to achieve full gain and PDE. Another important fact that can be identified in equation~\eqref{eq:SiPMVoltageRecovery} is the proportionality of the amplitude of the exponential functions to the released charge $Q$ and consequently the gain. A simulation thus becomes very simple. The released charge $Q$ is proportional to the voltage drop over the cell. Knowing only this voltage drop when the breakdown occurs is all that is needed to calculate the time evolution of all three voltages across the SiPM including the measured voltage pulse.

\subsection{Implementation of the simulation}
For the calculations performed in the previous paragraph only one cell is assumed to break down at time $t=0$. In a more realistic scenario, the SiPM is illuminated by a light pulse that is compiled from many photons. They impinge on the SiPM at arbitrary times and initiate multiple breakdowns of various cells. Due to the complexity of this scenario, an analytical calculation of the response of the SiPM is hardly feasible or not even possible at all. But, a careful investigation of equation~\eqref{eq:SiPMVoltageRecovery} allows drawing conclusions to perform a simulation of the SiPM response also for arbitrary time distributions of impinging photons.

Before any photon impinges on the SiPM, the applied overvoltage at all cells is $\Vbi$. When a photon impinges on a cell at time $t$, the released charge $Q\propto\Vbi$ is calculated in dependency of the applied overvoltage and released in a $\delta$-pulse. Knowing the exact time of the incident photon thus suffices to perform further calculations for the recharge of the cell. As all other variables in equation~\eqref{eq:SiPMVoltageRecovery} are constants, the time evolution of the deviation from $\Vbi$ of $\Vone$ and $\Vtwo$, $\tilde{\Vone}$ and $\tilde{\Vtwo}$ (cf.~eq.~\eqref{eq:SiPMVoltageRecovery}), respectively and $\Vthree$ can be determined directly. Then, for the time evolution of the voltage at the triggered cell, the voltage $\tilde{\Vone}$ is added and for all other cells the voltage $\tilde{\Vtwo}$ is added. The voltage $\Vthree$ contributes to the output voltage.

For a quantitative simulation, the correct dependencies of the gain and PDE on the overvoltage need to be known. For the gain, equation~\eqref{eq:SiPMGain} is well proven so that only the gain at one given overvoltage must be measured. For the PDE, a measurement of the relative change with the overvoltage is needed. For simplification, dead area and conversion efficiency are neglected which leads to a PDE of \SI{100}{\percent} at maximum overvoltage. Only the relative change of the PDE for not fully recovered cells is of interest in this case. While the measurement of the absolute PDE is complex because of the necessary knowledge of the absolute light flux incident on the SiPM, the relative change in PDE with the overvoltage can be obtained with low effort. A simple method that was used here is described in appendix~\ref{app:PDEMeasurement}.

\subsubsection{Workflow of the simulation}\label{sec:SimulationWorkflow}
For the full simulation of the SiPM, the incident photons impinge on random cells consecutively. For each photon, the current overvoltage of the hit cell is determined. Depending on the corresponding gain and PDE, a charge $Q$ is released and crosstalk photons are produced. The impact of the released charge on the voltages $\Vone$, $\Vtwo$ and $\Vthree$ is calculated. These steps will be described in detail in the following and the full workflow is sketched in figure~\ref{fig:SiPMSimulationFlowDiagram}. The pseudo code of the simulation is given in appendix~\ref{app:PseudoCode}.
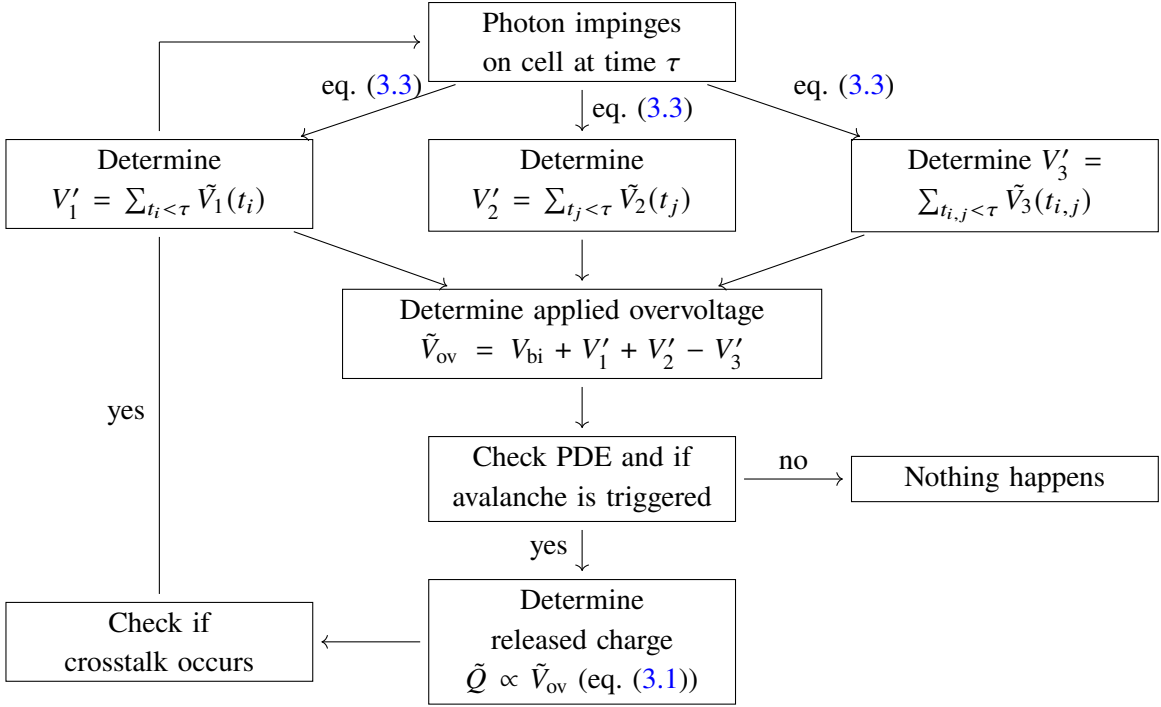
\begin{figure}\centering
    \input{img/SiPMSimulationWorkflow.tex}
    \caption{A schematic of the workflow of the simulation of the response of an SiPM for impinging photons. Details are given in the text.}
    \label{fig:SiPMSimulationFlowDiagram}
\end{figure}%

For each impinging photon, the currently applied voltage at the hit cell needs to be determined. Its value is influenced by three contributions:
\begin{enumerate}
    \item The state of the recharge of the same cell due to previous breakdowns $\Vone'(t)=\sum_{t_i}\tilde{\Vone}(t_i)$ where the sum goes over all previous photons that hit the same cell.
    \item The deviation from the nominal overvoltage due to previous breakdowns of other cells $\Vtwo'(t)=\sum_{t_j}\tilde{\Vtwo}(t_j)$ where the sum goes over all previous photons that hit other cells.
    \item The voltage at the readout $\Vthree'=\sum_{t_{i,j}}\Vthree(t_{i,j})$ where the sum goes over all previous photons.
\end{enumerate}
From these voltages, the instantaneous overvoltage $\Vovinst$ at a certain time $\tau$ can be determined as
\begin{equation}\label{eq:InstantaneousOvervoltage}
    \Vovinst(\tau) = \Vbi+\Vone'+\Vtwo'-\Vthree'= \Vbi+\sum_{t_i<\tau}\tilde{\Vone}(t_i)+\sum_{t_j<\tau}\tilde{\Vtwo}(t_j)-\sum_{t_{i,j}<\tau}\Vthree(t_{i,j})
\end{equation}
where $t_i$ refers to the times of all photons that hit the same cell earlier and $t_j$ refers to earlier hits of other cells.

From this instantaneous overvoltage, also the instantaneous PDE and gain $\tilde{g}$ can be determined. For the PDE, the curve shown in figure~\ref{fig:PDEMeasurement} in appendix~\ref{app:PDEMeasurement} is used. For the gain, the following equation is valid due to the proportionality with the overvoltage
\begin{equation}
    \tilde{g} = g\frac{\Vovinst}{\Vov}
\end{equation}
with $g$ being the gain when a voltage of $\Vov$ is applied. The value of $g$ needs to be known in advance. It can usually be taken from the datasheet and is given for a specific bias voltage. As will be seen in section~\ref{sec:IntrinsicParameterMeasurement}, the corresponding overvoltage should be measured individually.

With the probability describing the instantaneous PDE a random choice is made whether an avalanche is initiated. If this is not the case, no further action is performed as the photon is not detected. In case of an initiated avalanche, a charge $\tilde{Q}\propto\tilde{g}\propto\Vovinst$ is released and this needs to be taken into account when calculating $\Vovinst$ for the next photon according to equation~\eqref{eq:InstantaneousOvervoltage}.

Optical crosstalk can now easily be added on the base of a random process. For all SiPM cells being fully recovered, i.e.~all cells have a PDE of 1, the probability $\pemit$ for a photon to be emitted in the avalanche is directly given by the crosstalk probability $\pxt$. $\pemit$ scales proportional to the gain $\pemit=\pxt\cdot\tilde{g}/g$. Here, a random choice is made with probability $\pemit$ if a photon is emitted. If it is emitted, a randomly chosen directly neighboring cell is hit by this photon\footnote{Typically, crosstalk photons do not move far through the silicon or protective layer on top of it. A fraction of the crosstalk photons can hit also further distant cells due to internal reflections. Their contribution is low and can therefore be neglected here.}. The whole simulation is now performed for this cell to decide whether an avalanche is created, which charge is released and if again a photon is emitted from that cell. The procedure is repeated with probability $\pemit$ as long as a photon is emitted. The results obtained using this algorithm were tested to be identical to those presented in~\cite{FACTCalibration} where an extensive study of the crosstalk was performed. The algorithm presented here is simpler because it works correctly with the crosstalk probability $\pxt$ given in the datasheet while in~\cite{FACTCalibration} a correction must be applied.

Some features of SiPMs are not implemented for the purpose of simplification such as random fluctuations of the amount of released charge, thermal noise or afterpulsing. As we do not aim at giving reasonable results in the regime of only a few incident photons, these effects can be neglected. In a realistic SiPM, all signals propagate with finite speed. This effect is only relevant if a significant number of photons arrive within the average propagation time of the signals across the SiPM. As the propagation time is $\mathcal{O}(\SI{10}{\pico\second})$, even for \num{1e6} photons per \SI{100}{\nano\second} only about \num{100} photons arrive within this time range. The effect is thus negligible here and all signals are assumed to propagate instantaneously through the SiPM.

\section{Measurement of intrinsic parameters based on the simulation}\label{sec:IntrinsicParameterMeasurement}
Comparing the simulation to measurements, the intrinsic parameters of an SiPM can be determined. This will be done for the example of an SiPM of type Hamamatsu S13360-6025PE of size \SI{6x6}{\square\milli\meter} and a cell pitch of \SI{25}{\micro\meter} resulting in a total of \num{57600} cells.

The focus of this measurement is on the correct simulation of the exponential recharge of the SiPM cells. This is a crucial characteristic for the compensation of the non-linear response that will be discussed in section~\ref{sec:DynamicRangeExtension}. In order to do so, two consecutive light flashes illuminate the SiPM. The first flash initiates the breakdown of a fraction of the SiPM cells and the second pulse results in additional breakdowns. A measurement of the charge in the second pulse as a function of the delay between the two then allows to determine the state of the recharge of the cells.

\subsection{Measurement setup}
\begin{figure}\centering
    \includegraphics[width=\textwidth]{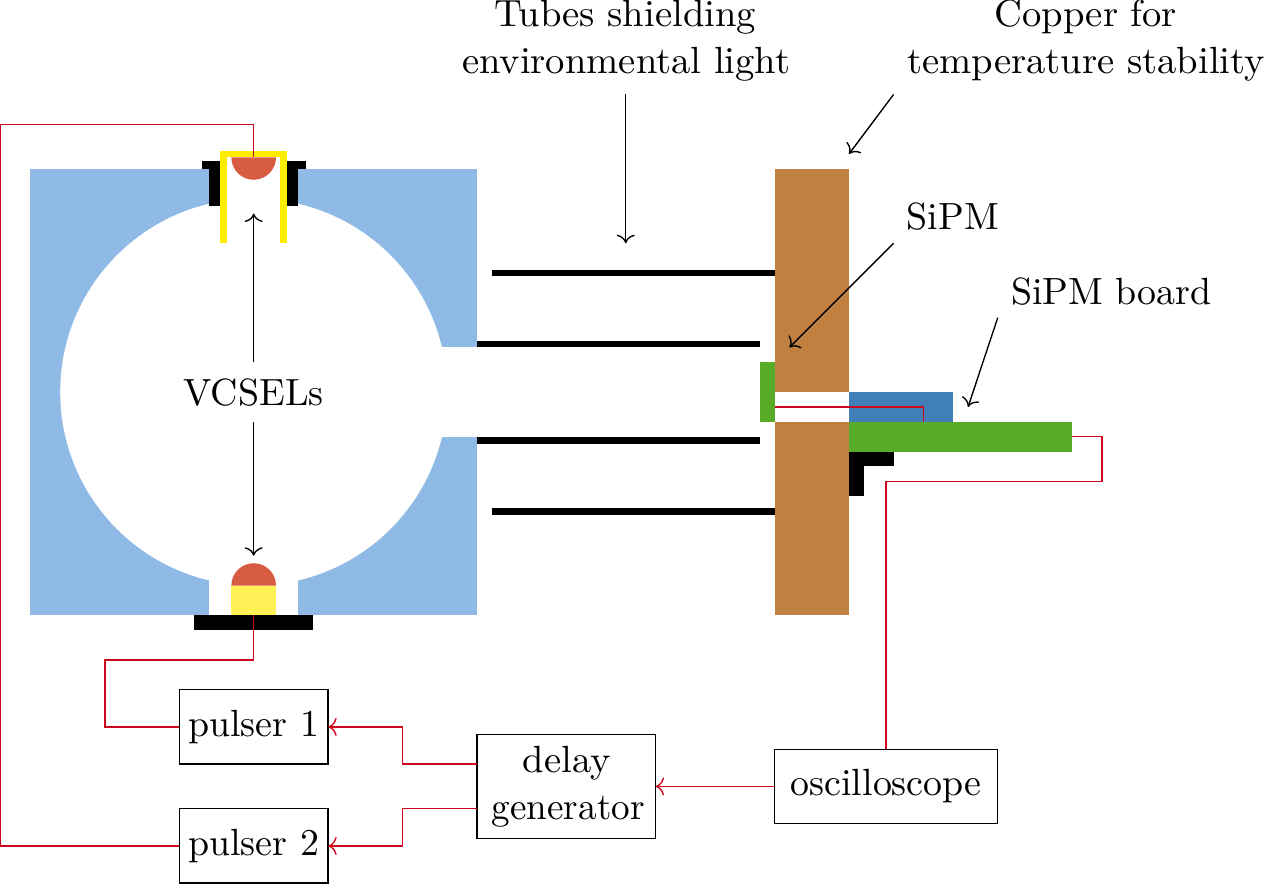}
    \caption{A sketch of the setup used for the determination of the intrinsic parameters of the SiPM. Two pulsers drive the VCSELs. Due to the slightly different versions of the pulsers also the mechanical layout of the VCSELs differs which is indicated in the sketch. The pulsers are triggered by a delay generator allowing for a variable delay between the two. An oscilloscope digitizes the SiPM signal and triggers the pulse creation. The SiPM is hosted on a copper block which is temperature stabilized. The SiPM board is used for mechanical stability of the SiPM positioning and for providing the SiPM signal at a standard LEMO connector. Taken from~\cite{ThesisJulian}.}
    \label{fig:IntrinsicParameterMeasurementSetup}
\end{figure}

The setup is sketched in figure~\ref{fig:IntrinsicParameterMeasurementSetup}. All details will be described in the following.
Three characteristics of the measurement setup are crucial for obtaining reliable results:
\begin{enumerate}
    \item The length of the light flashes should be at most the shortest intrinsic time constant of the SiPM to avoid consecutive hits of one cell. The exact values depend on the SiPM type but are typically at the order of only a few ns for the shortest time constant $\tau_+$.
    \item The pulses must be bright enough to trigger a significant fraction of the SiPM cells. If only a few cells are triggered, no cells are triggered twice and the effect will be insignificant.
    \item For a precise simulation, an exact knowledge of the spatial distribution of the light on the SiPM is required. In the most simple case, the SiPM is illuminated homogeneously which will also be done here.
\end{enumerate}
The first two items require a careful choice of the light source. Typical LEDs are very slow and cannot be brightly pulsed for only $\sim\SI{1}{\nano\second}$. Here, the choice fell on two picosecond light sources~\cite{mrongen}, referred to as \emph{pulsers} in the following, driving Vertical Cavity Surface Emitting Lasers (VCSELs) emitting light of wavelength \SI{850}{\nano\meter}. This configuration shows a width of only $\sim\SI{100}{\pico\second}$ and is therefore ideally suited for this measurement. Measurements of the electronic signal produced by the SiPM when being illuminated by one pulse have shown that a fraction of around \SI{50}{\percent} of the cells can be triggered in the final setup implying a sufficient brightness of the pulses. Two slightly different versions of the pulser were used. They differ in their mechanical layout and the used type of VCSEL. For one pulser, the VCSEL is a surface mounted device and soldered directly on the circuit board. For the second pulser, the VCSEL is connected to the pulser by a coaxial cable. The VCSELs of different mechanical layout also had different optical powers resulting in differently sized pulses. The second pulse is smaller by roughly one fourth. This does not affect the measurement.

A homogeneous light distribution is required on the SiPM. This is achieved by installing the two VCSELs in an integrating sphere. The reflections inside the sphere ensure a homogeneous distribution on the SiPM which is installed near one of its ports. The multiple reflections inside the sphere also cause a temporal broadening of the pulse. A toy Monte Carlo simulation tracking photons through the integrating sphere has been performed and reveals an exponential time distribution with characteristic time constant $\tau=\SI{3.44+-0.01}{\nano\second}$. The width of the optical pulse from the pulser is not included here but with only \SI{100}{\pico\second} it is negligible. The time $\tau$ is still short and well understood. It is included in the simulation of this measurement.

The SiPM is powered by a Hamamatsu C11204-02 integrated circuit which allows to set the bias voltage with a precision of \SI{10}{\milli\volt} and has negligible noise and temperature dependency~\cite{HamamatsuC11204}.

In order to achieve a precise measurement, the temperature needs to be kept stable. This is done by installing the SiPM on a copper block which is connected to piezo elements used for heating or cooling. The setup is surrounded by isolating material allowing precise control over the temperature. The deviation of the SiPM temperature from the nominal temperature is \SI{0.07}{\kelvin} at most. This device is described in detail in~\cite{ThesisCarsten}. All measurements were carried out at a temperature of \SI{25.0}{\degreeCelsius}. This is the temperature for which datasheet values of SiPM properties are given by the manufacturer.

The two pulsers are triggered by a delay generator. It allows to set variable delays between the two trigger lines with a precision of \SI{5}{\pico\second} and a jitter of less than \SI{100}{\pico\second}. Both effects are negligible compared to the length of the light pulse of \SI{3.44}{\nano\second} which was found earlier. The delay at the delay generator was corrected for additional constant delays that originate for example from different cable lengths. This constant delay was calibrated using two well separated pulses with delays $t_\mathrm{d}>\SI{500}{\nano\second}$. The true difference was measured in the recorded trace allowing the determination of the offset. For all measurements shown in the following this offset is corrected.

The oscilloscope provides a signal generator. It is used to trigger the delay generator and initiates the readout of the SiPM signal. Readout is performed at an impedance of $\Rs=\SI{50}{\ohm}$ without any pre-amplifiers to avoid distortions of the SiPM signal due to e.g.~limited bandwidth. This corresponds to the electronic circuit shown in figure~\ref{fig:SiPMElectricCircuit}. The signal is recorded with an analogue bandwidth of \SI{350}{\mega\hertz} at a sampling rate of 4\,Gs/s. A full trace has \num{20000} samples corresponding to a total time of \SI{5}{\micro\second}. The very long trace allows to fully record both pulses including the rather long tail even in the case of large delays.

\subsection{Measurement procedure}
\begin{figure}
    \centering
    \includegraphics[width=\textwidth]{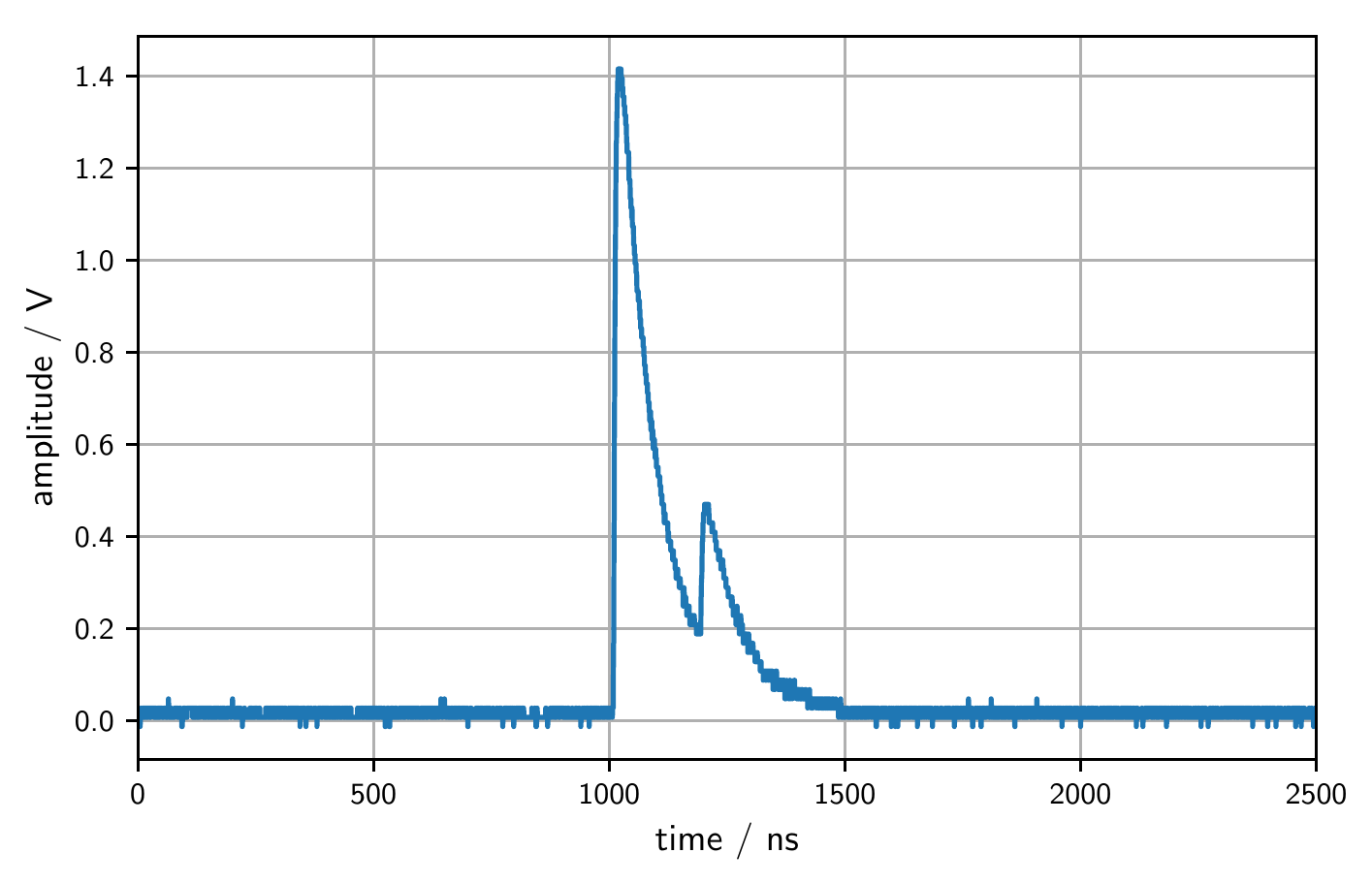}
    \caption{An example trace with two pulses with a delay of \SI{196}{\nano\second}. The trace is zoomed in, it extends up to \SI{5000}{\nano\second}. Taken from~\cite{ThesisJulian}.}
    \label{fig:DoublePulseTrace}
\end{figure}

The measurement is carried out in multiple steps. In order to allow a determination of the reduction of the second signal due to the first signal, its nominal size without the first pulse needs to be measured first. For each chosen delay, a total of three measurements is performed with \num{100} traces recorded:
\begin{enumerate}
    \item Only pulser 1 is switched on. A signal $S_1$ is measured at the SiPM.
    \item Only pulser 2 is switched on. A signal $S_2$ is measured at the SiPM.
    \item Both pulsers are switched on. A signal $\Sdel$ is measured at the SiPM.
\end{enumerate}
An example of a trace with two pulses is shown in figure~\ref{fig:DoublePulseTrace}.
The signals $S_1$, $S_2$ and $\Sdel$ are given by the sum over all samples in a time range from \SIrange{875}{5000}{\nano\second} subtracted by the pedestal which is the average amplitude in the first \SI{875}{\nano\second} of the trace. The long integration window is chosen to ensure that the full tail of the second pulse is included also in the case of large delays. The window also includes dark counts and afterpulses. Dark counts can be neglected as they are also contained in the pedestal region and thus corrected for. The amount of afterpulses scales with the initial signal size and they are contained in all three signals $S_1$, $S_2$ and $\Sdel$. In the analysis only relative changes of signals are compared and thus the contribution from afterpulses cancels out.

Due to the finite recharge times of the SiPM cells, $\Sdel<S_1+S_2$ is expected. This means that the sum of the signals of the two individual pulses alone is larger than the total signal when both illuminate the SiPM consecutively. The relative reduction of the second pulse due to the existence of the first pulse is
\begin{equation}\label{eq:R2}
    r_2 = \frac{\Sdel-S_1}{S_2}.
\end{equation}
The measurement is repeated for different delays in the range of \SIrange{-100}{2000}{\nano\second} where the largest delays are used for the calibration of the delay as was described previously. For negative delays, the order of the two pulses is inverted.

In addition, the measurement is carried out at different overvoltages. As gain, PDE and crosstalk have different dependencies on the overvoltage, the additional information helps to isolate different effects.

\subsection{Measurement results}
An example of a measured trace with two pulses with a delay of \SI{196}{\nano\second} was shown in figure~\ref{fig:DoublePulseTrace}. This allows to determine the relative reduction $r_2$ of the second pulse due to the first pulse according to equation~\eqref{eq:R2}.

\begin{figure}
    \centering
    \includegraphics[width=\textwidth]{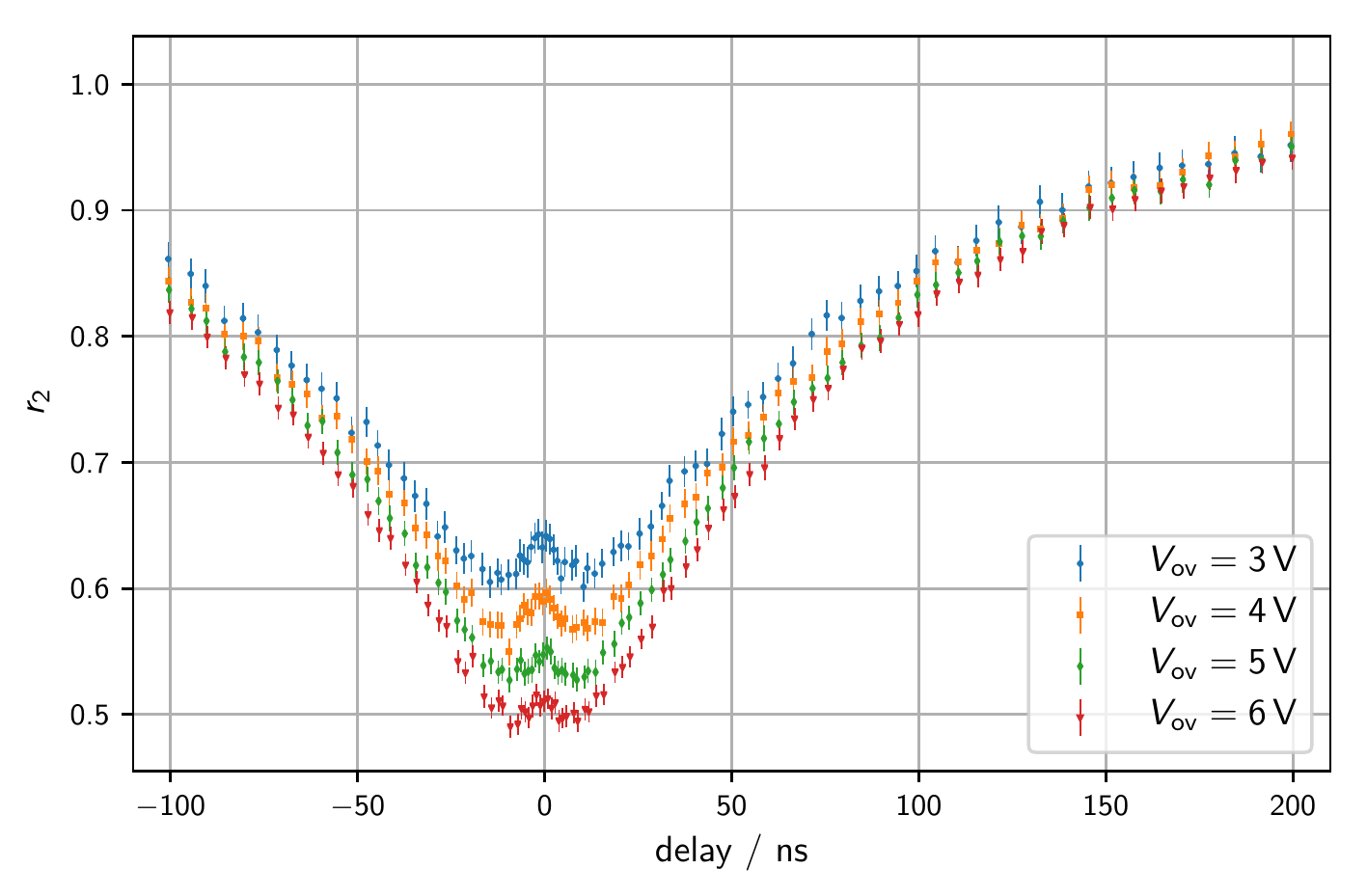}
    \caption{Comparison of the relative reduction of the second pulse due to the first one $r_2$ for different applied overvoltages. The effect increases with the overvoltage. The shown errorbars correspond to the uncertainty on the mean of 100 traces. Taken from~\cite{ThesisJulian}.}
    \label{fig:DoublePulseResult}
\end{figure}

The result is shown in figure~\ref{fig:DoublePulseResult} for different applied overvoltages.
As expected, the signal of the second pulse reduces significantly around a delay of \SI{0}{\nano\second}. The amount of reduction increases with the overvoltage and reaches up to \SI{50}{\percent} for $\Vov=\SI{6}{\volt}$.

The maximum reduction is not reached at a delay of \SI{0}{\nano\second} as might be naively expected. Instead, it is at delays of about \SI{10}{\nano\second}. In figure~\ref{fig:DoublePulseVoltageRecovery}, the time evolution of the voltage across the SiPM cells is depicted for the intrinsic values given by the manufacturer. The voltage drop over the untriggered cells (green dash-dotted line) reaches its minimum only about \SI{30}{\nano\second} after the avalanche was initiated. At this delay and not at \SI{0}{\nano\second}, the response of the untriggered cells is minimal causing the effect observed in the measurement. The slight asymmetry of the measured peak around zero in figure~\ref{fig:DoublePulseResult} originates from the differently sized pulses.

\subsection{Simulation of the measurement}
In order to determine the intrinsic parameters of the SiPM from this measurement, the simulation presented in section~\ref{sec:Simulation} is fit to the measurement result. To lower the computational cost of the simulation, a PDE of 1 at the nominal overvoltage is set. This avoids the simulation of photons that would not be detected anyway\footnote{It should be noted that this simplification has a slight impact on the fluctuations in the simulation. Due to the lower number of simulated photons also the absolute variation of the simulated signal size is lower than in the case of the correct PDE. This effect can be neglected here because of the large number of triggered avalanches.}.

The impinging photons from each of the two pulses are randomly distributed in time according to an exponential with time constant of $\tau=\SI{3.44}{\nano\second}$ as was simulated for the output of the integrating sphere. The photons are time ordered and hit random SiPM cells consecutively. For each photon, a specific charge according to the workflow in section~\ref{sec:SimulationWorkflow} is released. As for the measurement, the simulation is performed for each pulse individually to determine reference signals and also for the case of both pulsers being switched on.

In order to perform a proper simulation of the measurement, the number of impinging photons needs to be known. The signal $S$ in terms of p.e.~is given by:
\begin{equation}\label{eq:VoltagePEConversion}
    \frac{N}{\mathrm{p.e.}} = \frac{S}{\Rs f_\mathrm{s}}\frac{1}{e\cdot{}g}
\end{equation}
with $f_\mathrm{s}$ being the sampling rate and $e$ the elementary charge. The gain $g$ can be optimized in the fitting procedure starting from the datasheet value. The simulation needs to reproduce the same signal in terms of p.e. The number of impinging photons necessary to reproduce this signal is determined in an iterative process.

A total of six variables is optimized for the simulation to agree with the measurement. These are the intrinsic parameters from the electronic model $\Rq$, $\Cd$ and $\Cg$, the gain $g$, the crosstalk probability $\pxt$ and the breakdown voltage $\Vbd$. For the measurement, the breakdown voltage of $\Vbd=\SI{52.15}{\volt}$ given by the manufacturer was used as a reference to determine the overvoltage. In the fit, the breakdown voltage is a free parameter to account for a possible uncertainty. The capacitance of the quenching resistor $\Cq$ is not fit because it can be obtained from the other parameters using equation~\eqref{eq:SiPMGain}.

For each function evaluation in the fit procedure, two steps need to be performed:
\begin{enumerate}
    \item The number of impinging photons to reproduce the measured signal needs to be obtained for each of the two pulses and for each overvoltage individually.
    \item The obtained number of photons is put into the SiPM simulation for each given delay individually and the resulting $r_2$ is obtained.
\end{enumerate}
The measurement procedure is fast so that many data points could be taken. Implementing the fitting procedure, it was found that the function evaluation is slow and a speed up is necessary to obtain results within reasonable time. For each of the four overvoltages only every fourth data point is thus used in an alternating way. The simulation therefore has to be performed only on a fourth of the data points without loosing significant accuracy. A major complication in the fitting procedure is the stochastic nature of the simulation. It leads to fluctuations of the resulting function values and thus also to fluctuations in the $\chi^2$-value obtained for the agreement with the measurement.

\begin{figure}\centering
	\includegraphics[width=\textwidth]{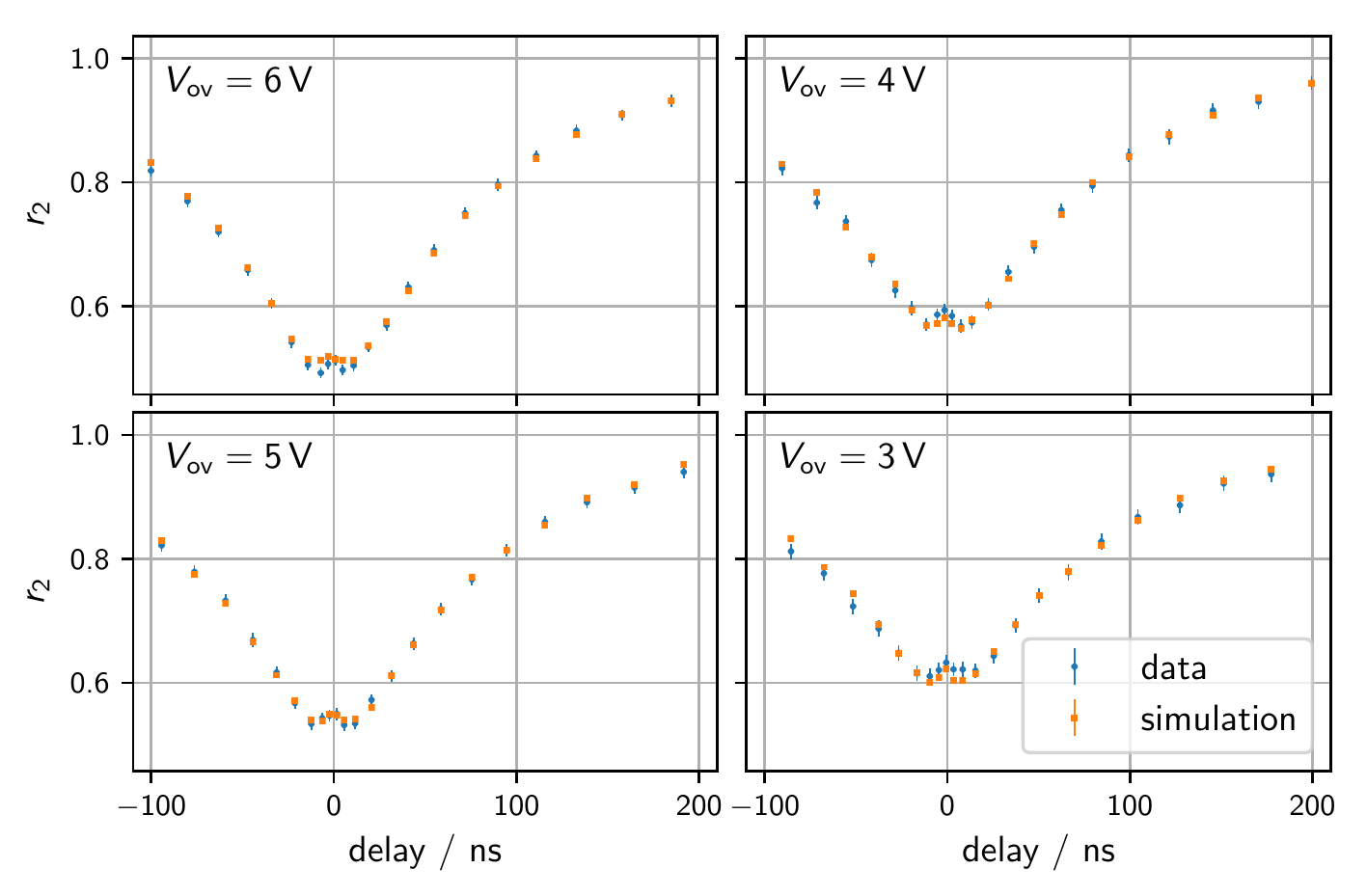}
	\caption{The relative size $r_2$ of the second pulse due to the first pulse as a function of the delay. Shown is the result of the fit given in the $\langle{}x\rangle$ row of table~\ref{tab:DoublePulseFitResults} together with the measurement. Both agree very well with a $\chi^2$ per degree of freedom of around 0.7. The indicated overvoltage is determined with respect to the breakdown voltage given by the manufacturer, not the fit result.}
	\label{fig:DoublePulseFitResult}
\end{figure}

The widely used gradient based fitting routines therefore do not succeed here. The simplex based Nelder-Mead method~\cite{NelderMead} was chosen instead in its implementation in scipy~\cite{Scipy}. The fit was repeated multiple times with different initial simplices. The resulting fit values are given in table~\ref{tab:DoublePulseFitResults} and the obtained agreement with the measurement is shown in figure~\ref{fig:DoublePulseFitResult}.
\begin{table}\centering
	\begin{tabular}{|c|c|c|c|c|c|c|}\hline
		fit & $\Rq$ / \SI{}{\kilo\ohm} & $\Cd$ / fF & $\Cg$ / pF & $g$ / $10^5$ & $\pxt$ / \% & $\Vbd$ / V\\\hline\hline
		1 & 844 & 20.6 & 38.6 & 6.93 & 5.9 & 51.68\\\hline
		2 & 785 & 20.8 & 42.1 & 7.18 & 6.9 & 51.84\\\hline
		3 & 789 & 20.8 & 40.2 & 7.28 & 4.7 & 51.91\\\hline
		4 & 796 & 20.8 & 40.9 & 7.37 & 4.0 & 51.97\\\hline
		5 & 758 & 21.2 & 43.7 & 6.90 & 3.9 & 51.99\\\hline
		$\langle{}x\rangle{}$ & $794\pm31$ & $20.84\pm0.22$ & $41.1\pm1.9$ & $7.13\pm0.21$ & $5.1\pm1.3$ & $51.914\pm0.066$\\\hline
		ref. & $790.4\pm6.4$ & $20.6\pm2.1$ & $41.1\pm4.1$ & 7 & 1 & $51.845\pm0.048$\\\hline
	\end{tabular}
	\caption{Resulting values from five fits of the SiPM simulation to the measurement with different initial simplices. The reference values in the last row are taken from the following sources: $\Rq$ and $\Vbd$ were determined in independent measurements for the identical SiPM. $\Cd$ and $\Cg$ are given by the manufacturer~\cite{HamamatsuPrivateCommunication} with an uncertainty of \SI{10}{\percent} and $g$ and $\pxt$ are given in the datasheet with unknown precision~\cite{HamamatsuS13360}. The second to last row indicates the mean and RMS of the fit results.}
	\label{tab:DoublePulseFitResults}
\end{table}
For each parameter the mean and RMS are calculated from the results of the five performed fits and given in the second to last row of table~\ref{tab:DoublePulseFitResults}. Typically, the uncertainty on a fit parameter is determined from its necessary variation to increase the $\chi^2$ value by 1. Due to the fluctuations of the simulation, this point cannot be determined here. Instead, the scatter of the fit results is a measure for the possible variations of the fit parameters and thus the RMS is taken as the uncertainty on the parameter.

The values given by the manufacturer or determined in independent measurements agree within the uncertainties with the results obtained in this measurement except for $\pxt$. This proves the good functioning of the described model. For $\Vbd$ a slight tension can be observed between the reference value and the measurement here but it is still acceptable within the uncertainties. For the crosstalk probability $\pxt$, the discrepancy between the measurement and the reference is significant. On the other hand, this measurement is likely not very sensitive to the crosstalk probability due to the small influence on the recharge behavior of the SiPM. Independent measurements also yield significantly deviating results from the reference, e.g.~in~\cite{ThesisJohannes} with $\pxt=\SI{7\pm0.2}{\percent}$. In addition, the exact optical setups can also modify the crosstalk probability meaning that different measurement methods might yield different results~\cite{BretzRehbeinCrosstalk}.

\subsection{Summary of the measurement results}
The simulation of an SiPM based on the equivalent electronic circuit presented in this section allows to describe the discussed measurement with two consecutive bright pulses with parameters being in the range expected from independent references. Performing measurements with two consecutive light flashes, each triggering a significant fraction of the SiPM cells, enables the determination of all intrinsic SiPM parameters of the model at once. Typically, each of them needs to be measured individually in dedicated setups. The recharge behavior of the SiPM is correctly modeled. As will be seen in the next section, the developed simulation allows to make predictions also for other types of measurements especially in the case of very bright pulses.

\section{Extending the linear dynamic range of SiPMs}\label{sec:DynamicRangeExtension}
As has been seen in the previous section, the SiPM signal does not scale linearly with the number of incident photons. This is a result of the limited number of cells, the finite recharge time after a breakdown occurred and also the impact of the current flowing through the other cells after a breakdown. The dynamic range of an SiPM is therefore intrinsically non-linear and a complex function of the number of incident photons and also their time distribution.

For simultaneously impinging photons, the output signal $\Nmeas$ in terms of p.e.~is the result of a binomial process and, neglecting crosstalk and afterpulsing, can be written as~\cite{RenkerSiPMOverview}:
\begin{equation}\label{eq:SiPMDynamicRange}
    \Nmeas = \Ncell\left(1-e^{-\textit{PDE}\cdot N_{\mathrm{inc}}/\Ncell}\right)\,\mathrm{p.e.}
\end{equation}
with $\Ncell$ the number of cells and $N_{\mathrm{inc}}$ the number of incident photons.

Though the response is non-linear, a linear approximation is feasible in the regime where $N_\mathrm{inc}\ll\Ncell$ holds. For instance, for the Hamamatsu S13360-6025PE used in the previous section with \num{57600} cells and a PDE of around \SI{25}{\percent}, a deviation from linearity of \SI{10}{\percent} is reached for $N_\mathrm{inc}\sim\num{50000}$. A total of \num{11000} cells gets triggered in that case of which roughly \SI{10}{\percent} get hit by two or more photons. This is only a fourth of the number of cells while the rest of the SiPM does not actively contribute to the measurement. Given the fact that this SiPM is among those with the largest number of cells available, the dynamic range of a single SiPM is limited to around \num{10000} photons without correcting for the non-linearity.

Equation \eqref{eq:SiPMDynamicRange} is only valid for simultaneously impinging photons. For pulses extended in time, the response curve is more complex and cannot be calculated analytically. The number of incident photons cannot easily be determined from the measured signal. Here, an algorithm has been developed to correct for the non-linearity also when the time distribution of the impinging photons is not known. It exploits the full measured voltage trace which contains the time information and makes use of the simulation that has been introduced in the previous section.

\subsection{The algorithm to extend the linear dynamic range}\label{sec:DynamicRangeExtensionAlgorithm}
In order to reconstruct the incident number of photons without prior knowledge of the time distribution of the incident photons, the measured voltage trace needs to be exploited. It contains temporal information, but the photon arrival time distribution is smeared by the response of the SiPM. The focus of this algorithm is to reconstruct the total number of impinging photons, not their time distribution. This is nonetheless done by finding the time distribution of incident photons $\NgammaOft$ that reproduces the measured voltage signal $\SOft$ best when being detected by the SiPM.

This could simply be done in a brute force approach by guessing the incident photon arrival time distribution. Comparing the resulting voltage signal with the measurement and adjusting the photon time distribution accordingly, the correct incident photon distribution could be reconstructed in an iterative process. Many iterations are necessary in this case and finding the correct adjustment of the photon distribution after each iteration is difficult due to the broad smearing of the voltage signal.

\begin{figure}
    \centering
    \includegraphics[width=\textwidth]{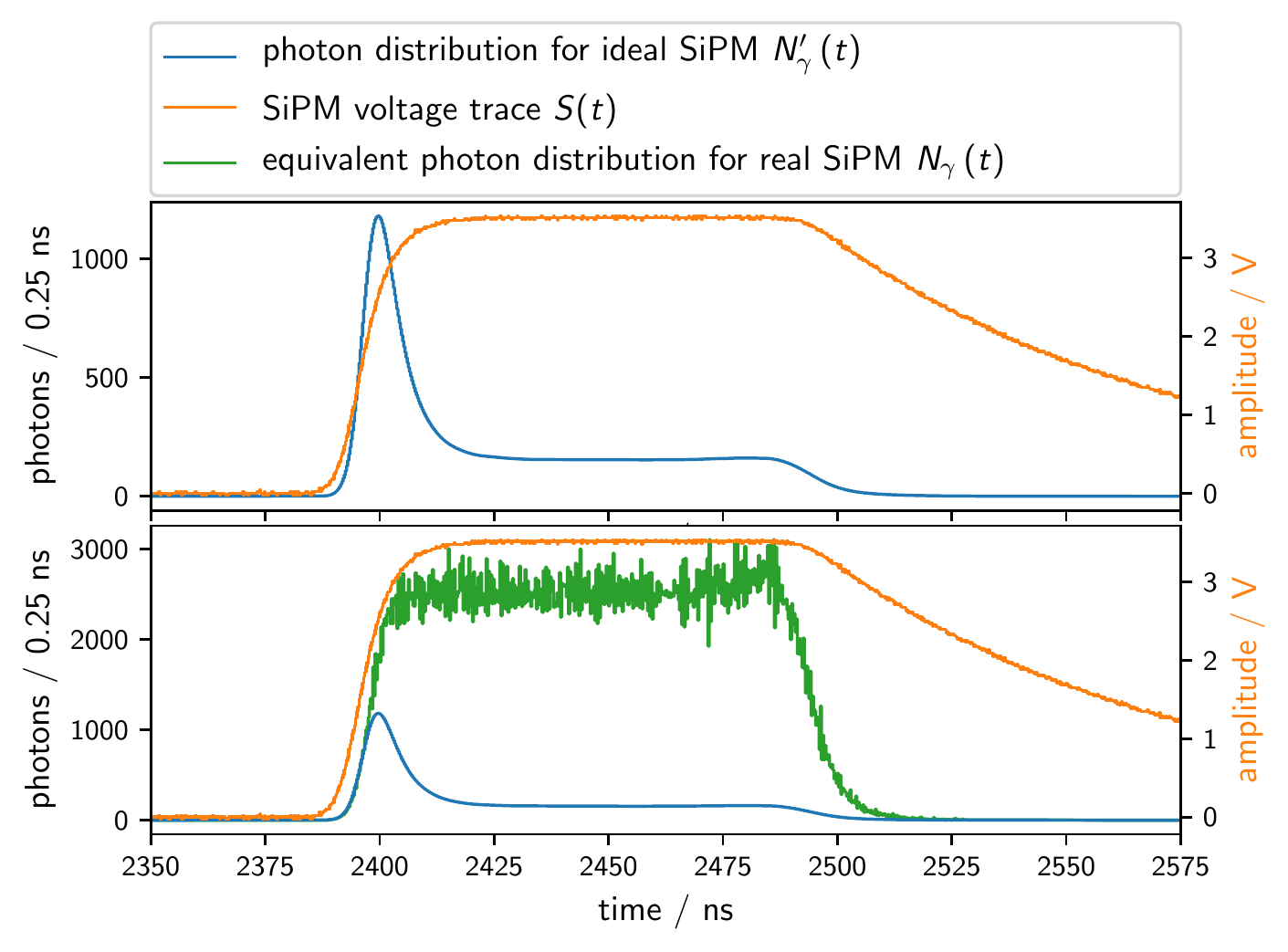}
    \caption{The two steps for the reconstruction of the incident equivalent photon time distribution from a measured voltage signal $\SOft$. Note the different y-scales.}
    \label{fig:ReconstructionSteps}
\end{figure}

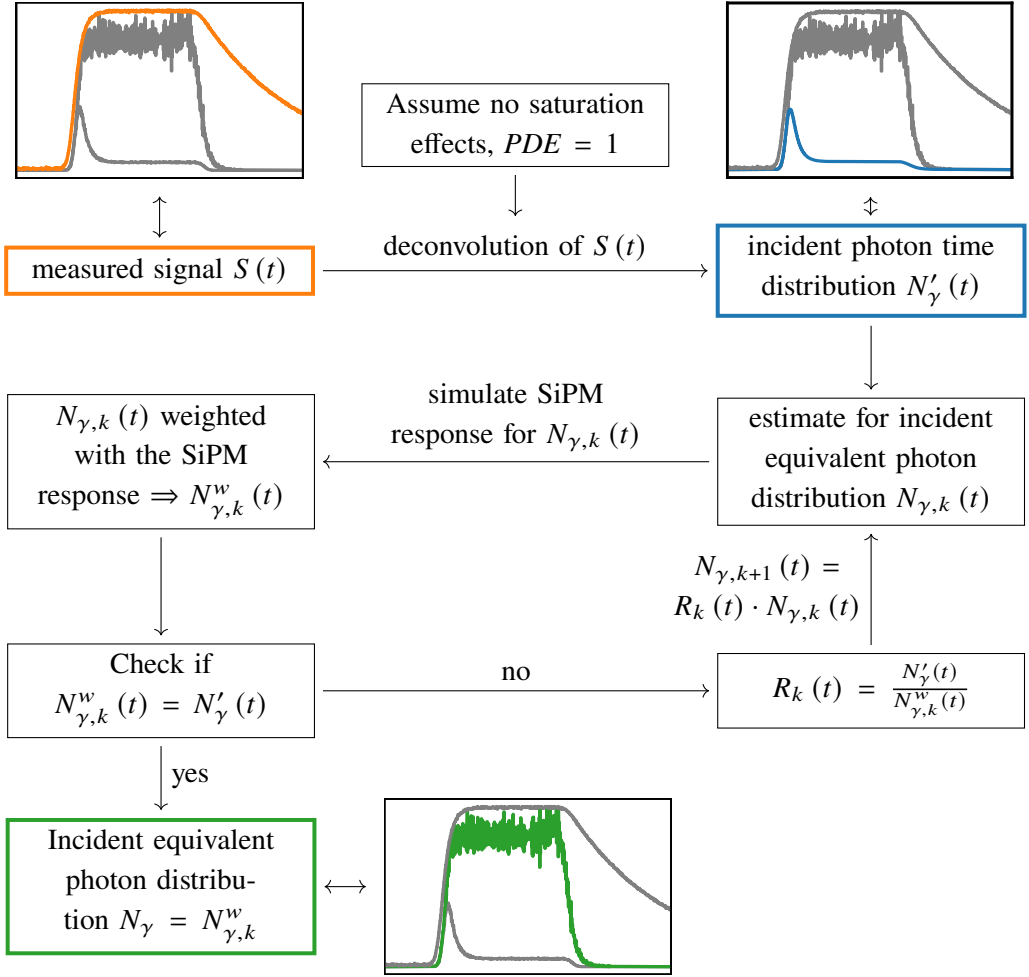
\begin{figure}
    \centering
    \input{img/SiPMDeconvolutionSchematic.tex}
    \caption{A schematic of the algorithm used to extend the linear dynamic range of SiPMs. The graphs in the inlets and their colors are identical to those in figure~\ref{fig:ReconstructionSteps}. Details are given in the text.}
    \label{fig:ReconstructionAlgorithm}
\end{figure}

A better approach was found by performing two separate steps:
\begin{enumerate}
    \item An ideal SiPM is assumed. It does not suffer from any saturation effects and each incident photon initiates an avalanche at a fully recovered cell\footnote{This corresponds to a PDE of \SI{100}{\percent} and is used for simplification here. If one is interested in the real number of incident photons, the obtained result needs to be divided by the true PDE.}. The time distribution of incident photons $\NgammaPrimeOft$ that reproduces the measured signal $\SOft$ is calculated for this sensor. This is sketched in the top of figure~\ref{fig:ReconstructionSteps}. The procedure will be described in detail in section~\ref{sec:IdealSiPMIncidentPhotonDistribution}.
    \item The incident equivalent photon time distribution $\NgammaOft$ for the real SiPM is determined. The term \emph{equivalent} refers to the fact that a PDE of \SI{100}{\percent} is assumed so that the distribution obtained here needs to be divided by the true PDE if one is interested in the true incident photon distribution. The term \emph{real} for the SiPM describes the fact that the full recharge behavior of the cells is included in contrast to the \emph{ideal} SiPM where each incident photon initiates an avalanche at a fully recovered cell. The signal $f$ in terms of p.e.~produced by each incident photon can be obtained from the SiPM simulation. Then, the correct $\NgammaOft$ is given if it reproduces $\NgammaPrimeOft$ using the individual $f$ as weights. In this case, the incident equivalent photon time distribution $\NgammaOft$ will produce the measured voltage signal $\SOft$ for the real SiPM. This is sketched in the bottom of figure~\ref{fig:ReconstructionSteps}. The procedure will be described in section~\ref{sec:RealSiPMIncidentPhotonDistribution}.
\end{enumerate}
In the following paragraphs, the algorithm will be explained in detail by guiding through the sketch given in figure~\ref{fig:ReconstructionAlgorithm}.
A list of the symbols used in the following sections is given in appendix~\ref{app:ListOfSymbols}.

\subsubsection{Determination of the incident photon distribution for an ideal SiPM}\label{sec:IdealSiPMIncidentPhotonDistribution}
Here, an ideal SiPM is assumed where each incident photon triggers an avalanche at a fully recovered cell. In this case, the measured voltage signal $\SOft$ is given by the convolution of the incident photon distribution $\NgammaPrimeOft$ with the single p.e.~voltage signal $\Vspe\cdot\SPEOft$. Here, $\Vspe$ and $\SPEOft$ denote the amplitude and shape, respectively, of the signal. Additional electronics noise $\noiseOft$ is added in typical applications:
\begin{equation}\label{eq:IdealSiPMConvolution}
    \SOft = \Vspe\cdot\NgammaPrimeOft * \SPEOft + \noiseOft = \Vspe\cdot\int_{-\infty}^\infty N'_\gamma (t')\cdot\textit{SPE}(t-t')\mathrm{d}t' + \noiseOft .
\end{equation}
For the case of the SiPM, the shape of $\SPEOft$ can either be measured or approximated by an exponential rise and fall. Here, the functional form
\begin{equation}
    \SPEOft = c\cdot\frac{1}{1+e^{-t/\tau_+}}\cdot{}e^{-t/\tau_-}
\end{equation}
is used with $\tau_\pm$ from equation~\eqref{eq:SiPMTimeConstants} and $c$ denoting a normalization constant.

Obtaining $\NgammaPrimeOft$ from equation~\eqref{eq:IdealSiPMConvolution} corresponds to a deconvolution which is a widely discussed topic in the literature~\cite{ValinoICRC2015,ImageDeconvolution}. Various algorithms were developed for specific tasks and optimized to deal with different issues. The major difficulty is the presence of the usually unknown noise term $\noiseOft$ which can have a significant impact on the deconvolution.

Here, an algorithm based on~\cite{PyUnfold,DAgostiniUnfolding} was found to achieve very good results. It is based on Bayes' theorem and was originally developed for counting experiments. Though the problem presented here is not a counting experiment, it can mathematically be described similarly as will be seen in the following.

In order to make use of Bayes' theorem, the function $\SPEOftAndTau$ should be seen as the probability density function for an incident photon at time $\tau$ to produce a voltage signal at a time $t$. For the rest of this section, $t$ will always refer to the time of the measured voltage signal and $\tau$ to the time of an incident photon. When neglecting noise and assuming a discrete sampling of the voltage signal $\SOft$, equation~\eqref{eq:IdealSiPMConvolution} can be written as a matrix multiplication:
\begin{equation}\label{eq:IdealSiPMMatrixSolution}
    \SOft = \Vspe\cdot{}M\cdot\NgammaPrimeOftau
\end{equation}
with $M_{ij}=\textit{SPE}(t_i|\tau_j)$ being the response matrix of the ideal SiPM. Neglecting the additional term $\noiseOft$ is justified here because of the focus on large signals where the signal-to-noise ratio is high. Nonetheless, $\noiseOft$ is still present in $\SOft$ and can significantly spoil the result when simply solving equation~\eqref{eq:IdealSiPMMatrixSolution} by inverting $M$. The noise might be amplified when being divided by small values of~$M$.

Instead, a pseudo-inverse of $M$ will be determined using Bayes' theorem. It allows to avoid the division by small values. The probability $\textit{SPE}'(\tau_j|t_i)$ that a photon hit the SiPM at time $\tau_j$ under the assumption of a specific measured voltage signal $S(t_i)$ at time $t_i$ is given by
\begin{equation}\label{eq:SPEPrime}
    \textit{SPE}'(\tau_j|t_i) = \frac{\textit{SPE}(t_i|\tau_j)P(\tau_j)}{\sum_k \textit{SPE}(t_i|\tau_k)P(\tau_k)}
\end{equation}
with $P(\tau_j)$ being the probability for a photon to hit the SiPM at time $\tau_j$. From this equation, the incident photon distribution for the ideal SiPM $N'_\gamma{}(\tau_i)$ can then simply be obtained from
\begin{equation}
    N'_\gamma{}(\tau_i) = \frac{1}{\Vspe}\sum_j \textit{SPE}'(\tau_i|t_j)S(t_j).
\end{equation}

The only remaining problem is the unknown probability $P(\tau_i)$. For a given incident photon distribution $N'_\gamma{}(\tau_i)$, it can be calculated as
\begin{equation}
    P(\tau_i) = \frac{N'_\gamma{}(\tau_i)}{\sum_j N'_\gamma{}(\tau_j)}.
\end{equation}
The lack of knowledge can be compensated by starting from a prior distribution $P_0(\tau_i)$ and using an iterative approach. First, the corresponding function $\textit{SPE}'_0(\tau_j|t_i)$ is determined using equation~\eqref{eq:SPEPrime}. This leads to a first estimate $N'_{\gamma,0}(\tau_i)$ on $N'_\gamma{}(\tau_i)$. Using an initially uniform prior yields:
\begin{align}
\begin{split}
    P_0(\tau_i) &= \frac{1}{N_\mathrm{samples}}\\
     \textit{SPE}'_k(\tau_j|t_i) &= \frac{\textit{SPE}(t_i|\tau_j)P_k(\tau_j)}{\sum_m \textit{SPE}(t_i|\tau_m)P_k(\tau_m)}\\
     N'_{\gamma{},k}(\tau_i) &= \frac{1}{\Vspe}\sum_j \textit{SPE}'_k(\tau_i|t_j)S(t_j)\\
     P_{k+1}(\tau_i) &= \frac{N'_{\gamma,k}(\tau_i)}{\sum_j N'_{\gamma,k}(\tau_j)}
\end{split}
\end{align}
with $N_\mathrm{samples}$ being the number of samples of the trace and the index $k$ denoting the number of iterations performed. The iteration needs to be stopped, when a satisfying result is obtained. Then, the incident photon distribution for the ideal SiPM is found:
\begin{equation}
    N'_\gamma{}(\tau) = N'_{\gamma,k}(\tau).
\end{equation} 
As the major goal is a good agreement with the measured voltage trace $\SOft$, the reconstructed trace $S_k(t)$ is calculated after each iteration using equation~\eqref{eq:IdealSiPMMatrixSolution}:
\begin{equation}
    S_k(t) = \Vspe\cdot{}M\cdot{}N'_{\gamma,k}(t).
\end{equation}
The agreement of $S_k(t)$ with $\SOft$ can now, for example, be determined using a $\chi^2$-test. For the results shown in the following, the $\chi^2$-test is restricted to signal regions which are $3\sigma$ above the noise level. Based on a trial-and-error approach criteria were chosen for stopping the iteration. The first possibility is a $\chi^2$ per degrees of freedom below 2. While this precision is typically reached within a few seconds on a standard desktop PC, conversion of the algorithm gets slow and achieving better agreement takes several minutes. This value was thus chosen as a compromise. Alternatively, the iteration is stopped if the $\chi^2$ per degrees of freedom only improves by less than \SI{1}{\permil} in one iteration, again to avoid long computing times of even several ten minutes. For the presented results, this only happened in a few percent of the cases.

\begin{figure}
    \centering
    \includegraphics[width=\textwidth]{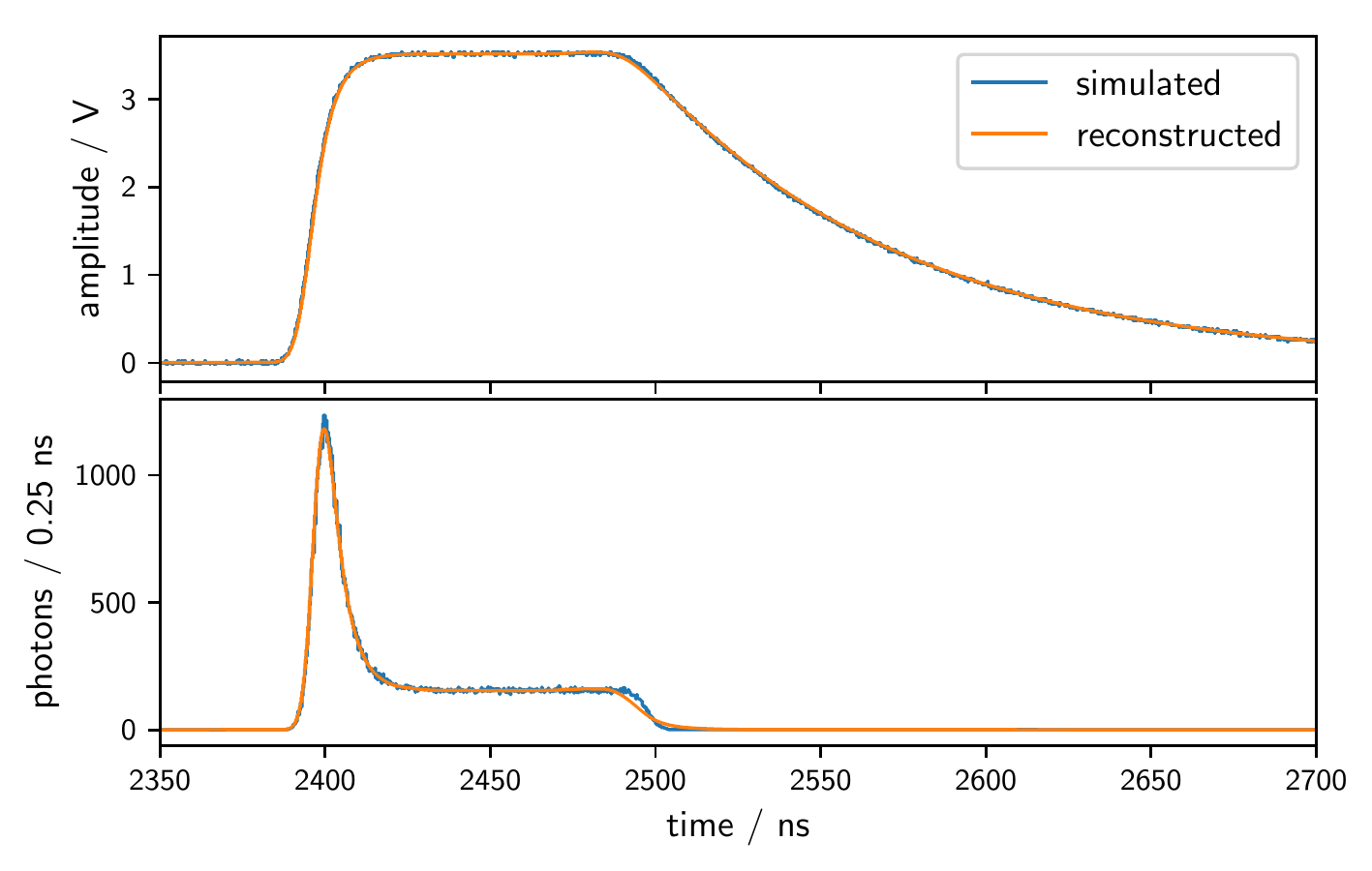}
    \caption{Comparison of a simulated and reconstructed trace for a simulation with $10^6$ incident photons within \SI{100}{\nano\second} and being Gaussian smeared by additional \SI{3}{\nano\second} for an SiPM of type Hamamatsu S13360-6025PE as used in section~\ref{sec:IntrinsicParameterMeasurement}. \emph{Top:} The originally simulated and the reconstructed traces $\SOft$ are in very good agreement. \emph{Bottom:} The simulated and reconstructed photon distributions $\NgammaPrimeOft$ under the assumption of an ideal SiPM where each photon triggers an avalanche. Taken from~\cite{ThesisJulian}.} 
    \label{fig:TraceReconstructionResult}
\end{figure}

In figure~\ref{fig:TraceReconstructionResult}, the result is shown for a simulation of $10^6$ incident photons within \SI{100}{\nano\second} being smeared with an additional Gaussian of \SI{3}{\nano\second} width.
The simulated and reconstructed traces agree within the noise level. The same holds for the photon distributions $\NgammaPrimeOft$. Only a slight mismatch is visible in the falling tail and around the peak of the distribution. As these differences are small, the reconstruction can be considered successful.

\subsubsection{Reconstruction of the incident equivalent photon time distribution for a real SiPM}\label{sec:RealSiPMIncidentPhotonDistribution}
In the previous section, an ideal SiPM was assumed in order to find an incident photon distribution $\NgammaPrimeOft$ that produces an output of $\SOft$ of the SiPM. For the transition to a realistic SiPM, all details of its response including the recharge behavior need to be taken into account. The main difference between the ideal and the real SiPM is the released charge for an impinging photon. For the ideal SiPM, the output always corresponds to the single p.e.~pulse $\SPEOft$ but for the real SiPM it is scaled the released charge $f_i$ in terms of p.e.:
\begin{equation}
    \SPERealOft = f_i\cdot\SPEOft.
\end{equation}
Here, $f_i$ is between 0 and 1 in all cases where a cell is not fully recharged but can also be larger than 1 in case of crosstalk. The goal of the algorithm presented in this section is to find the incident equivalent photon time distribution for the real SiPM $\NgammaOftau$ that equals $\NgammaPrimeOftau$ for the ideal SiPM when each photon is weighted with the corresponding $f_i$.

An iterative approach will be used to find $\NgammaOftau$. First, photons according to the incident photon distribution for the ideal SiPM $\NgammaPrimeOftau$ are put into the SiPM simulation introduced in section~\ref{sec:Simulation} and an $f_i$ is obtained for each photon. It allows to determine the photon distribution $N^\mathrm{w}_{\gamma,0}(\tau)$ when each photon is weighted with the released charge $f_i$.  $N^\mathrm{w}_{\gamma,0}(\tau)$ will usually be smaller than $\NgammaPrimeOftau$ due to $f_i<1$ in most cases. The time dependent ratio $R_0(\tau)=\NgammaPrimeOftau/N^\mathrm{w}_{\gamma,0}(\tau)$ can be calculated and its multiplication with the initial incident photon distribution for the ideal SiPM $\NgammaPrimeOftau$ corresponds to a new incident equivalent photon time distribution for the real SiPM $N_{\gamma,1}(\tau)$. Iterating this process yields:
\begin{align}
\begin{split}
    N_{\gamma,0}(\tau_i) &= N'_\gamma(\tau_i)\\
    N^\mathrm{w}_{\gamma,k}(\tau_i) &= \textit{SiPM}(N_{\gamma,k}(\tau_i))\\
    R_k(\tau_i) &= \frac{N'_\gamma(\tau_i)}{N^\mathrm{w}_{\gamma,k}(\tau_i)}\\
    N_{\gamma,k+1}(\tau_i) &= R_k(\tau_i)\cdot N_{\gamma,k}(\tau_i)
\end{split}
\end{align}
with $\textit{SiPM}(N_{\gamma,k}(\tau))$ being the charge distribution resulting from the SiPM simulation when using $N_{\gamma,k}(\tau)$ as input and the index $k$ referring to the number of iterations performed. For a reasonable agreement between $N^\mathrm{w}_{\gamma,k}(\tau)$ and $\NgammaPrimeOftau$ the iteration needs to be stopped. Then, the incident equivalent photon time distribution for the real SiPM is given by:
\begin{equation}
    \NgammaOftau = N_{\gamma,k}(\tau)
\end{equation}
and the total reconstructed number of photons that impinged on the SiPM is obtained from the integral of $\NgammaOftau$.

For the results presented in the following sections, the integrals of $N^\mathrm{w}_{\gamma,k}(\tau)$ and $N'_\gamma(\tau)$ are compared with the mean fraction of released charge per photon $\bar{f}$:
\begin{equation}
	\frac{\int_{-\infty}^\infty N'_\upgamma\left(\tau\right)\text{d}\tau
	- \int_{-\infty}^\infty N_{\upgamma,k}^\text{w}\left(\tau\right)\text{d}\tau}{\int_{-\infty}^\infty N'_\upgamma\left(\tau\right)\text{d}\tau} < 0.005\cdot\bar{f}.
\end{equation}
The dependency on $\bar{f}$ ensures a similar precision independent of the photon flux.

\begin{figure}
    \centering
    \includegraphics[width=\textwidth]{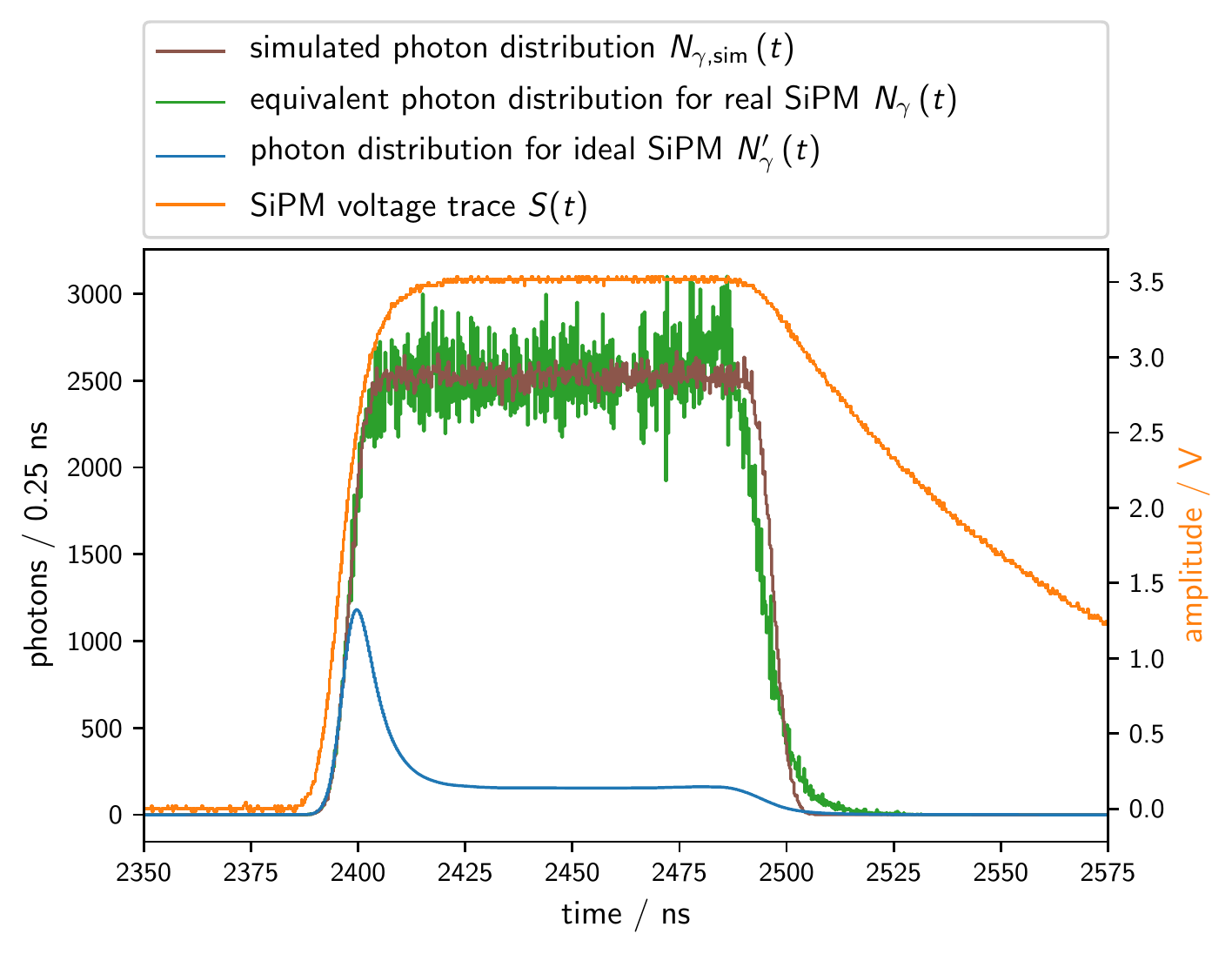}
    \caption{Comparison of the originally simulated and the reconstructed photon distribution for the same pulse as shown in figure~\ref{fig:TraceReconstructionResult}. The two agree within their noise level. Only in the falling tail a slight mismatch is visible. The distribution when taking the SiPM response into account as $\delta$-functions differs significantly meaning that a large correction needed to be applied. This also results in the larger fluctuations of the reconstructed distribution compared to the simulated one. In grey, the simulated SiPM voltage trace is shown for comparison.}
    \label{fig:IncidentPhotonDistributionReconstruction}
\end{figure}

An example for such a reconstruction applied on simulations is shown in figure~\ref{fig:IncidentPhotonDistributionReconstruction}. 
The incident photon distribution for the ideal SiPM $N'_\gamma(t)$ is significantly lower than the originally simulated incident equivalent photon time distribution $N_{\gamma,\mathrm{sim}}(t)$. This is the effect of the necessary recharge of the SiPM cells. After applying the presented algorithm, the agreement improves. As mentioned earlier, the focus was placed on the precise reconstruction of the amount of incident photons, not on their time distribution. With a total deviation of the total number of reconstructed photons by only \SI{0.6}{\percent} compared to the \num{1e6} simulated photons this goal is achieved. The rising edges of the reconstructed and simulated distributions, $\NgammaOft$ and $N_{\gamma,\mathrm{sim}}(t)$, respectively, agree with each other. In the plateau region, statistical fluctuations are larger for the reconstruction. Huge correction factors around 10 need to be applied to the distribution for the ideal SiPM $\NgammaPrimeOft$ and result in an amplification of the noise. The time distributions deviate in the falling tail where the reconstructed one falls slower compared to the simulation. As the main goal was not the reconstruction of the time distribution but the total number of reconstructed photons, a detailed study of this deviation was not performed here.

\subsection{Application to measurements}\label{sec:DynamicRangeMeasurement}
In order to verify the reconstruction algorithm, dedicated measurements were performed to study the dynamic range of SiPMs. The identical SiPM of type Hamamatsu S13360-6025PE as in section~\ref{sec:IntrinsicParameterMeasurement} was used, in the same setup that was introduced in figure~\ref{fig:IntrinsicParameterMeasurementSetup}. A few modifications had to be made to allow the measurement over the required wide dynamic range from single p.e.~level to high non-linearity:
\begin{enumerate}
    \item For the previously used VCSELs, the brightness and length of the pulse could not be varied. One of the VCSELs was thus replaced by a LED with a wavelength of \SI{390}{\nano\meter} and the corresponding pulser was replaced with a different one that allowed for a variation of the length of the electronic pulse up to \SI{100}{\nano\second} and also the amplitude. These components allow to perform measurements over a wide range in brightness.
    \item A reference for the brightness of the light pulse is needed. The second VCSEL was thus replaced by a photodiode of type Hamamatsu S2281-01~\cite{HamamatsuS2281}. It was connected to a picoammeter to measure its current $I_\mathrm{d}$. The relative brightness $S$ of a single pulse is then given by $S=I_\mathrm{d}/R_\mathrm{p}$ with $R_\mathrm{p}$ being the repetition rate of the pulser.
\end{enumerate}
The conversion of the current of the photodiode to an amount of photons impinging on the SiPM was calibrated with the SiPM signal in the range of quasi-linear response. The signal measured at the SiPM in terms of p.e.~$\Nmeas$ is given by equation~\eqref{eq:VoltagePEConversion}.

The repetition rate $R_\mathrm{p}$ of the LED was varied between \SI{20}{\hertz} and \SI{300}{\kilo\hertz} depending on the set brightness of the LED. For a very dim LED, the current at the photodiode gets low and a higher rate is advantageous while for a bright LED the SiPM gets saturated and consecutive pulses might influence each other. In that case, a lower rate is favorable.

\begin{figure}
    \centering
    \includegraphics[width=\textwidth]{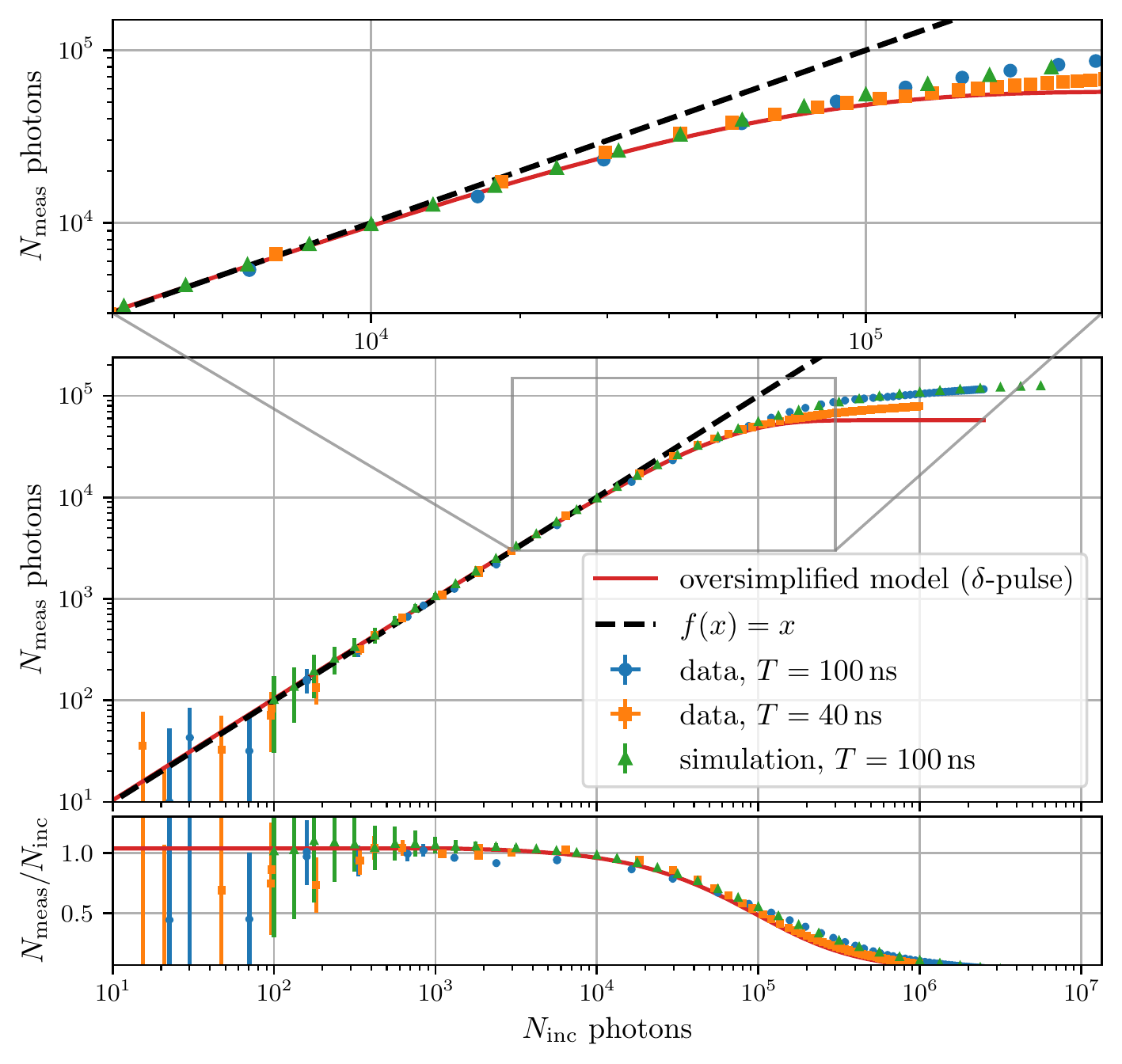}
    \caption{The response curve for the SiPM for two measurements with an incident pulse with \SI{40}{\nano\second} and \SI{100}{\nano\second} lengths. At the top, a zoom into the region where the deviation from linearity becomes significant is shown. For the shorter pulse, a significant deviation from linearity sets in earlier. In addition, the oversimplified model according to equation~\eqref{eq:SiPMDynamicRange} and a simulation with a \SI{100}{\nano\second} long pulse are shown.}
    \label{fig:SiPMResponseCurve}
\end{figure}

The resulting response curve for the SiPM is shown in figure~\ref{fig:SiPMResponseCurve} for two measurements with a \SI{40}{\nano\second} and \SI{100}{\nano\second} long pulse, a simulation with a \SI{100}{\nano\second} long pulse and the theory curve for an incident $\delta$-pulse of photons according to equation~\ref{eq:SiPMDynamicRange}.
The simulation is in good agreement with the measurement of the same pulse width. For a very high number of incident photons, the output signal exceeds the number of cells of $\Ncell=\num{57600}$. This effect only occurs in the case of non-simultaneously impinging photons. It is thus expected here and sometimes referred to as \emph{oversaturation}~\cite{Gruber:2013jia}. Due to the long duration of the light pulses, the SiPM cells can get triggered multiple times resulting in a sum that is larger than the number of cells. 

\begin{figure}
    \centering
    \includegraphics[width=\textwidth]{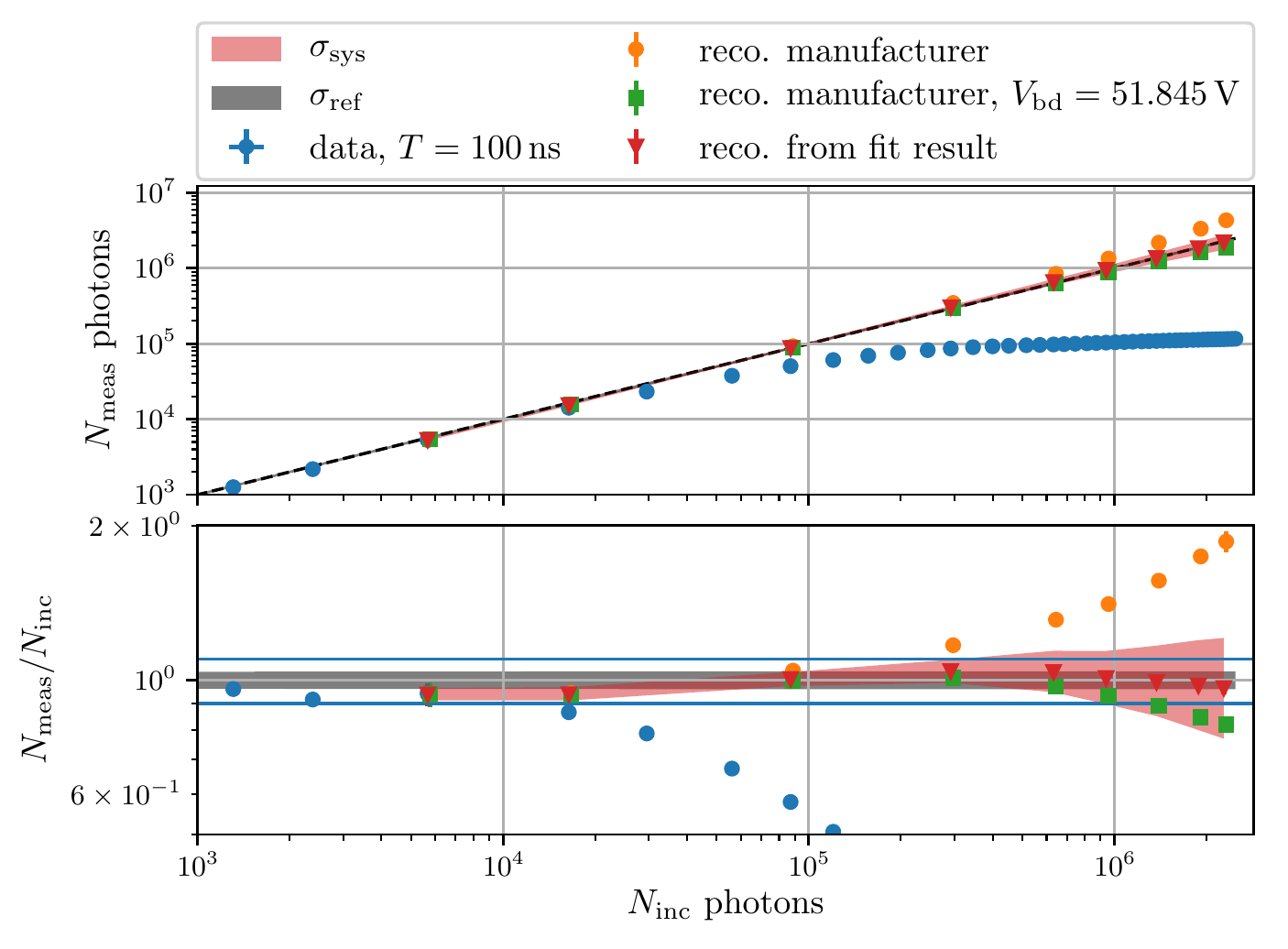}
    \caption{The reconstructed response curve of the SiPM for the measurement with a pulse length of \SI{100}{\nano\second}. The reconstruction was performed multiple times with different sets of variables. For the \emph{reco.~from fit result}, the result $\langle{}x\rangle$ from the fitting procedure presented in table~\ref{tab:DoublePulseFitResults} in section~\ref{sec:IntrinsicParameterMeasurement} was used. For the \emph{manufacturer}, the values of the manufacturer taken with and without the offset in the breakdown voltage that was found in the fit was used. The black shaded area corresponds to the uncertainty on the reference number of photons due to the calibration of the photodiode. The red shaded area indicates the systematic uncertainty originating from the \emph{fit results}. The blue horizontal lines correspond to \SI{10}{\percent} deviation from linearity. Modified from~\cite{ThesisJulian}.}
    \label{fig:ReconstructionResult100ns}
\end{figure}

\begin{figure}
    \centering
    \includegraphics[width=\textwidth]{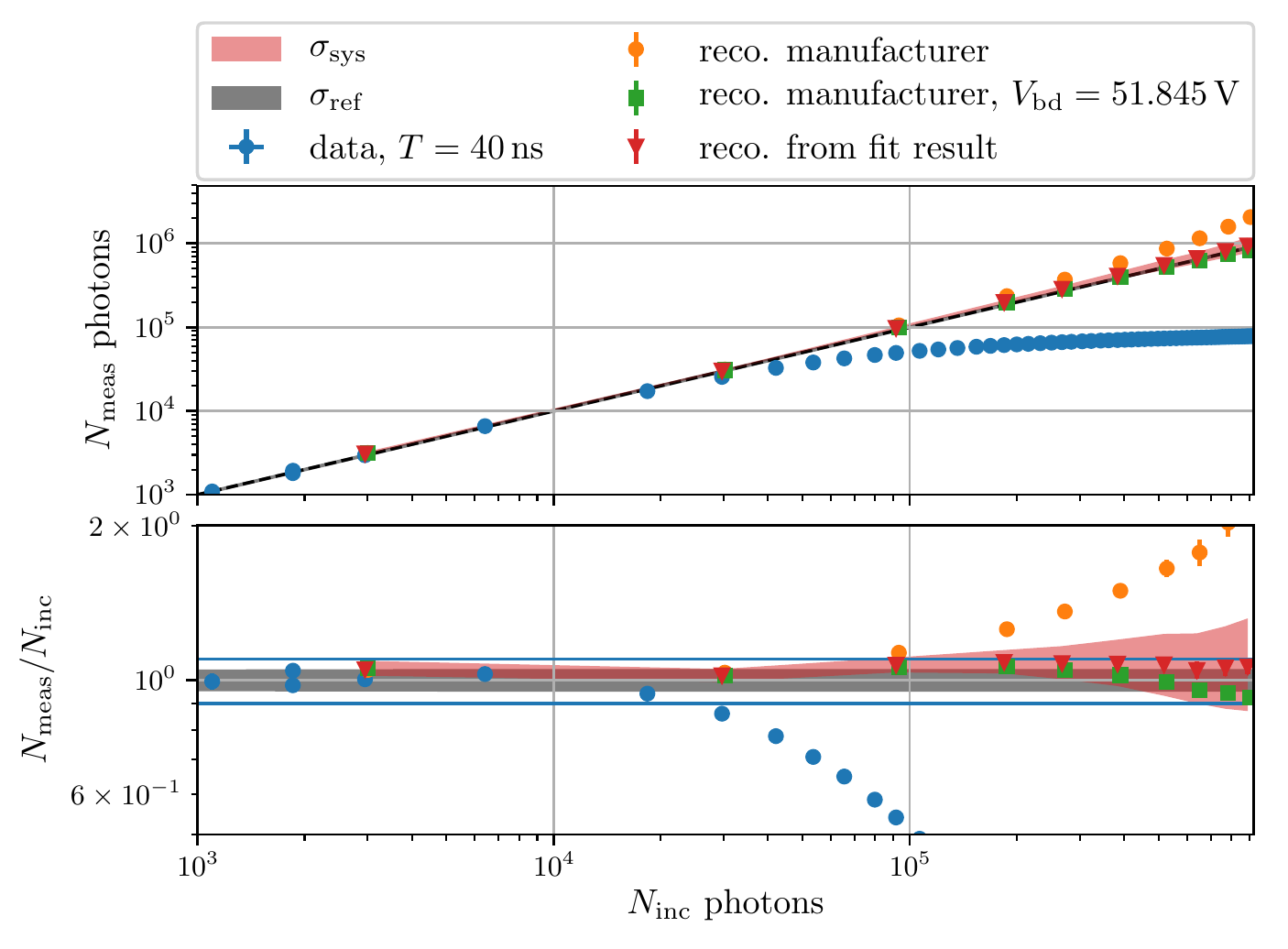}
    \caption{The same as in figure~\ref{fig:ReconstructionResult100ns} for a pulse width of \SI{40}{\nano\second}. Modified from~\cite{ThesisJulian}.}
    \label{fig:ReconstructionResult40ns}
\end{figure}

The algorithm presented in section~\ref{sec:DynamicRangeExtensionAlgorithm} is applied to the measurement to reconstruct the number of impinging photons. The results for the incident pulses with a length of \SI{100}{\nano\second} and \SI{40}{\nano\second} are shown in figures~\ref{fig:ReconstructionResult100ns} and~\ref{fig:ReconstructionResult40ns}.
In the reconstruction procedure, the SiPM simulation is a crucial part. The parameters of the simulation (cf.~tab.~\ref{tab:DoublePulseFitResults}) can be varied to understand their impact on the reconstruction result and the necessary precision of their knowledge. Three different sets of variables were used here:
\begin{enumerate}
    \item The result from the fit described in section~\ref{sec:IntrinsicParameterMeasurement} and given in the second to last row of table~\ref{tab:DoublePulseFitResults}.
    \item The intrinsic values for the used type of SiPM given by the manufacturer in table~\ref{tab:6025PEIntrinsicParameters}. It allows to study the performance of the algorithm in case no dedicated measurement of the intrinsic parameters of the specific used SiPM is carried out.
    \item Same as 2.~but the overvoltage is taken from an independent measurement of the current-voltage-curve. This value is given in the last row of the last column of table~\ref{tab:DoublePulseFitResults}. As the measurement of the breakdown voltage and consequently the overvoltage is relatively simple it can be carried out for individual SiPMs and will add valuable information to the reconstruction procedure.
\end{enumerate}
These three models are compared with each other by the increase in dynamic range for a deviation of~\SI{10}{\percent} from linearity. To achieve a linear reconstruction better than \SI{10}{\percent} the prevision of the setup needs to be improved.

The results from the fit performed in section~\ref{sec:IntrinsicParameterMeasurement} also yield the best result in the reconstruction. The reconstructed number of photons equals the incident number of photons within \SI{10}{\percent} in the studied dynamic range up to \num{2e6} incident photons. For the raw data this deviation is already reached at only \num{1e4}. With the presented algorithm, the dynamic range is improved by more than two orders of magnitude.

For the reconstructed response curve, a slight excess of up to \SI{5}{\percent} is found in the region between \num{e5} and \num{e6} incident photons for the \SI{100}{\nano\second} long pulse, and for a higher number of incident photons the reconstructed number becomes too low. For the shorter pulse of \SI{40}{\nano\second} the excess reaches \SI{9}{\percent} at \SI{2e5} incident photons. These excesses might be due to the uncertainties on the parameters resulting from the fit described in section~\ref{sec:IntrinsicParameterMeasurement} and given in the $\langle{}x\rangle$ row of table~\ref{tab:DoublePulseFitResults}.

\begin{figure}
    \centering
    \includegraphics[width=\textwidth]{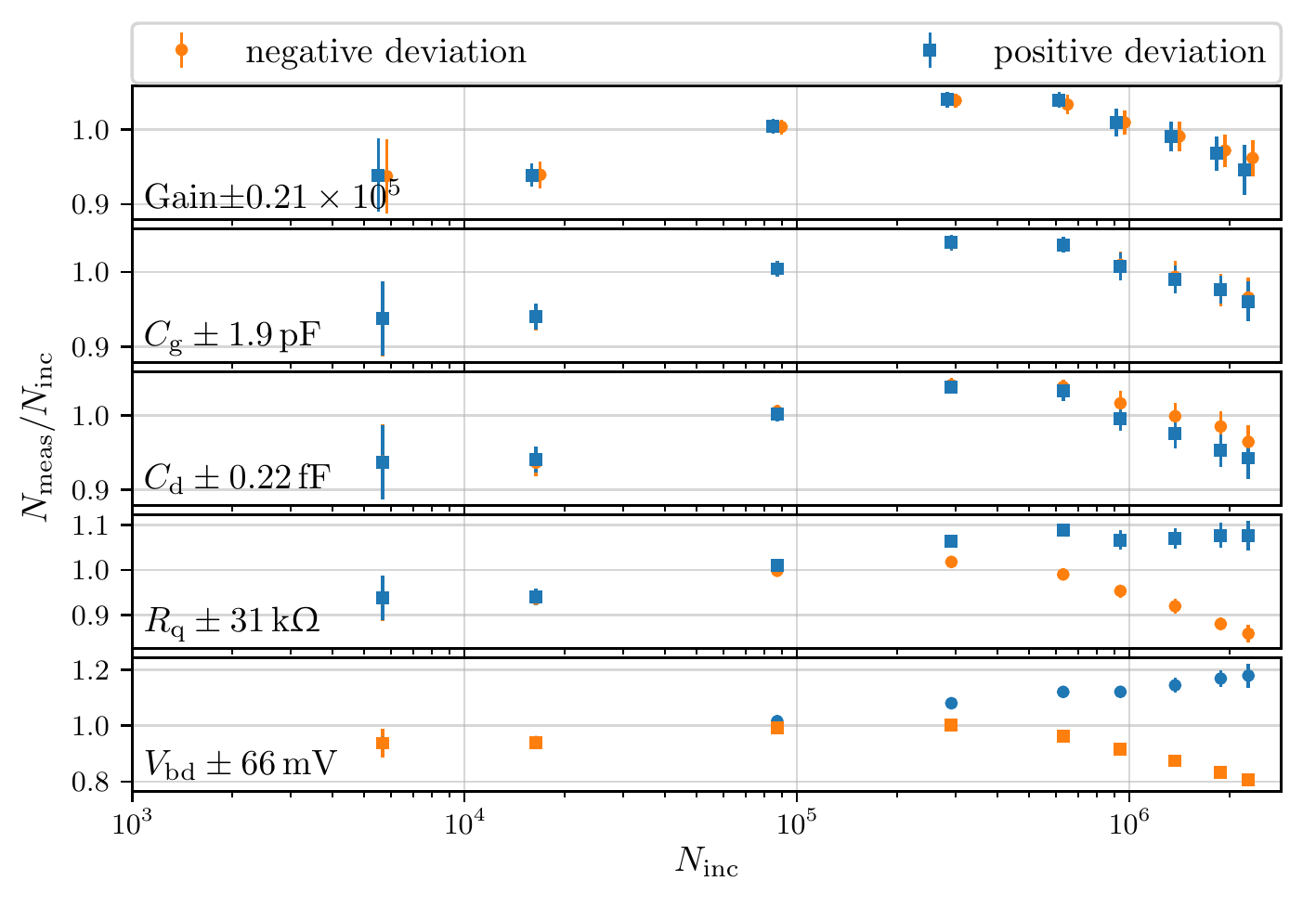}
    \caption{The impact of the uncertainties of the intrinsic SiPM parameters on the reconstruction. Each parameter is varied by its uncertainty according to the measurement shown in section~\ref{sec:IntrinsicParameterMeasurement}. The \emph{positive} (\emph{negative}) deviation corresponds to the results obtained when changing the corresponding parameter by one standard deviation in positive (negative) direction. Note the different y-scales. Modified from~\cite{ThesisJulian}.}
    \label{fig:ReconstructionResult100nsUncertainty}
\end{figure}

The systematic uncertainty on the reconstructed number of incident photons originating from the uncertainty on the parameters of the fit result is also shown in the figure. The reconstruction was repeated for each parameter being varied according to its uncertainty as given in row $\langle{}x\rangle{}$ of table~\ref{tab:DoublePulseFitResults}. How the resulting reconstructed number of incident photons consequently changes is depicted in figure~\ref{fig:ReconstructionResult100nsUncertainty} exemplarily for the pulse of width \SI{100}{\nano\second}. The same analysis was also performed for the \SI{40}{\nano\second} pulse and yields similar results.
The shaded red bands in figures~\ref{fig:ReconstructionResult100ns} and \ref{fig:ReconstructionResult40ns} correspond to the quadratic sum of the variations of the reconstructed number of incident photons obtained for the five different parameters in this analysis. This assumes statistical independence of the parameters which must not be true. For instance, gain, overvoltage and diode capacitance are correlated as can be seen from equation~\eqref{eq:SiPMGain}. Due to the complexity of this analysis, a determination of the correlation of the parameters was not possible.

It can clearly be seen that the uncertainties on the quenching resistor and on the breakdown voltage have the largest impact on the result. The uncertainty increases with the number of incident photons from \SI{3}{\percent} at $N_\mathrm{inc}=\num{1e5}$ and \SI{12}{\percent} at $N_\mathrm{inc}=\num{1e6}$ to \SI{20}{\percent} at $N_\mathrm{inc}=\num{2e6}$ for the \SI{100}{\nano\second} wide pulse. For the shorter pulse, the systematic uncertainty is \SI{4}{\percent} at $N_\mathrm{inc}=\num{1e5}$ and reaches \SI{25}{\percent} at $N_\mathrm{inc}=\num{1e6}$ incident photons. The uncertainties for the short pulse are larger by up to a factor of 2 at the same number of incident photons. This is a result of the higher number of photons per time for the short pulse compared to the long one.

Using the data provided by the manufacturer (cf.~tab.~\ref{tab:6025PEIntrinsicParameters} on page~\pageref{tab:6025PEIntrinsicParameters}) results in a worse performance. The point of \SI{10}{\percent} deviation from linearity is reached already at \num{1e5} incident photons. This corresponds still to an improvement by one order of magnitude compared to the raw data. Adding only the information on the exact breakdown voltage improves the result significantly and extends linearity by almost another order of magnitude. These observations strongly indicate that a precise knowledge of the intrinsic parameters is not necessary for the algorithm to work properly. For measurements where multiple SiPMs of the same type are used, this means that only the simple measurement of the breakdown voltage is necessary while for all other parameters, i.\,e.~the quenching resistance $\Rq$, quenching capacitance $\Cq$, diode capacitance $\Cd$, grid capacitance $\Cg$, crosstalk probability $\pxt$ and gain $g$, the values given by the manufacturer can be used. As can be seen in figure~\ref{fig:ReconstructionResult100nsUncertainty}, for further improvement of the result the quenching resistance needs to be known as it has the second largest impact on the result.

It should be noted that some systematic uncertainties have not been studied here. As the total resistance of the SiPM is only around \SI{10}{\ohm}, already small additional resistors in the circuit can influence the results. This might e.\,g.~be due to bad cable connections. Another source of uncertainty is the shape of the single p.e.~pulse which might be distorted due to the readout electronics. This can affect especially the very fast leading edge which could be broadened by too slow electronics.

\section{Conclusions}
A simulation of the response of an SiPM based on an equivalent electronic model was implemented and studied in a dedicated measurement setup. Measuring the response for two consecutive light flashes and adapting the model to the observables allowed to measure the intrinsic parameters of the SiPM and to correctly describe its recharge behavior. The results are in agreement with independent measurements.

The simulation was used to extend the linear dynamic range of SiPMs. The raw SiPM response is intrinsically non-linear and is given by the limited number of cells and their necessity of recharging after a breakdown occurred. An algorithm was developed that makes use of the full measured voltage trace to analyse the time development of the signal. It allows to reconstruct the number of incident photons with good precision. Measurements with a rectangular photon arrival time distribution revealed an improvement in the linearity of the response by at least two orders of magnitude compared to the raw data. This region of linearity even exceeds the number of cells of the SiPM by more than one order of magnitude. This improvement makes SiPMs ideally suited for many classes of experiments where a precise calibration of the single p.e.~signal is necessary and at the same time a wide dynamic range needs to be achieved. Similar improvements can be expected for different shapes of the photon arrival time distribution as the developed algorithm does not make any assumptions on it.

Future studies should focus on different readout schematics, environmental conditions and the variations between different SiPMs of the same type. The study of the systematic uncertainty of the reconstruction suggests that the latter can be neglected if an increase in linearity by around one order of magnitude is needed. The necessary precision of the data acquisition was not studied in the scope of this work. With a lower sampling rate, lower analogue bandwidth or worse resolution of the data acquisition than used in this work, the uncertainties on the reconstructed result might increase. Different readout schematics, e.\,g.~with additional amplifiers, lead to a different SiPM response. These changes can be included in the SiPM simulation so that the reconstruction is still possible.

Especially the environmental temperature can have an impact on the values of the intrinsic parameters, in particular the quenching resistance. While for old devices the quenching resistor was made of poly-silicon, recent devices have metal quenching resistors. This change improves the temperature dependency of the resistance by 1/5 down to \SI{0.2}{\percent\per\degreeCelsius} in recent devices~\cite{HamamatsuMPPCHandbook,Otte:2016aaw}. For temperature variations of up to \SI{\pm20}{\degreeCelsius} the change is within the uncertainty for the quenching resistor that was studied here.

\section{Acknowledgements}
We thank Martin Rongen for providing the pulsers and their layout which allowed to perform the double pulse measurements. We acknowledge the help of the electronic workshop of Physics Institute IIIA of RWTH Aachen University who designed the pulser for the dynamic range measurement. We thank Johannes Schumacher for his support during the initial phase of these measurements and simulations. Simulations were performed with computing resources granted by RWTH Aachen University under project rwth0364.

\bibliographystyle{JHEP}
\bibliography{main}

\appendix

\section{Differential equations for the SiPM equivalent electronic circuit}\label{app:DifferentialEquations}
The various currents and voltages in the circuit shown in figure~\ref{fig:SiPMElectricCircuit} form a set of differential equations:
\begin{align}
\begin{split}
    I_\mathrm{t}(t) &= \left(\Vtwo(t)-\Vbi\right)\frac{N-1}{\Rq}+\left(N-1\right)\Cq\frac{\mathrm{d}}{\mathrm{d}t}\left(\Vtwo(t)-\Vbi\right)\\
    I_\mathrm{q}(t) &= \left(\Vbi-\Vtwo(t)\right)\frac{1}{\Rq}+\Cq\frac{\mathrm{d}}{\mathrm{d}t}\left(\Vbi-\Vtwo(t)\right)\\
    I_\mathrm{g}(t) &= \Cg\frac{\mathrm{d}}{\mathrm{d}t}\left(\Vthree(t)-\Vbi\right)\\
    I_\mathrm{p}(t) &= \Cd\frac{\mathrm{d}}{\mathrm{d}t}\left(\Vthree(t)-\Vone(t)\right)\\
    I_\mathrm{u}(t) &= \left(N-1\right)\Cd\frac{\mathrm{d}}{\mathrm{d}t}\left(\Vthree(t)-\Vtwo(t)\right)\\
    I_\mathrm{SiPM}(t) &= \frac{\Vthree(t)}{\Rs}.
\end{split}
\end{align}
The currents can be related to each other by applying Kirchhoff's circuit laws. The resulting set of equations can then be solved analytically. Details can be found in~\cite{MaranoSiPMAnalyticalAnalysis,Jha:2013,MasterThesisJohannes,ThesisJulian}.

\section{Measurement of the relative PDE-overvoltage curve}\label{app:PDEMeasurement}
The measurement of the absolute PDE of an SiPM is rather complicated. A precise knowledge of the absolute light flux on the SiPM is necessary. The determination of an absolute light flux can be done with a calibrated photodiode. For conversion to a flux on the SiPM the exact geometries of the setup (such as spatial light distribution, positioning of the photodiode, positioning of the SiPM) need to be known, which is difficult to achieve. For the simulations and measurements presented in this publication, the absolute PDE does not need to be known. Instead, the relative change of the PDE with respect to the overvoltage is enough. A simple measurement method is presented in the following.

The average measured integrated signal $S(\Vov)$ for a given light pulse is directly proportional to the product of PDE and gain $g$ as long as the light flux is well below saturation of the SiPM:
\begin{equation}\label{eq:SignalGainPDE}
    S(\Vov)\propto g(\Vov)\cdot\textit{PDE}(\Vov).
\end{equation}
Here, contributions from crosstalk and afterpulsing are neglected which is a reasonable approximation due to their low contribution of only a few percent for the studied SiPM of type Hamamatsu S13360-6025PE~\cite{HamamatsuS13360}. According to equation~\eqref{eq:SiPMGain}, the gain is proportional to the overvoltage. Then, equation~\eqref{eq:SignalGainPDE} can be rewritten as
\begin{equation}\label{eq:PDESignalOV}
    \textit{PDE}(\Vov)\propto\frac{S(\Vov)}{g(\Vov)}\propto\frac{S(\Vov)}{\Vov}.
\end{equation}
Using a pulsed light source, the signal $S(\Vov)$ can be measured for different overvoltages. 

\begin{figure}\centering
    \includegraphics[width=\textwidth]{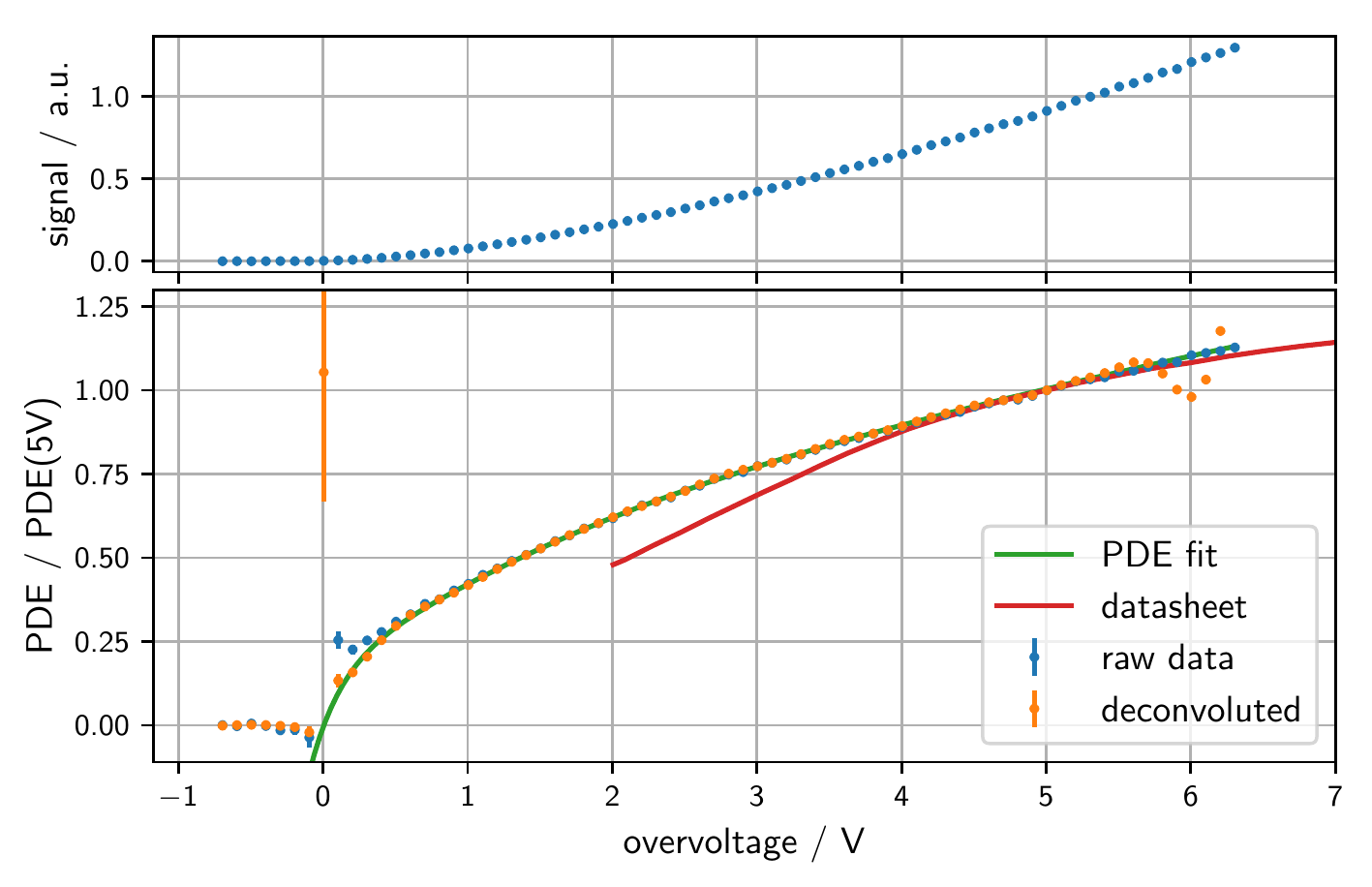}
    \caption{Measurement of the relative PDE with respect to the overvoltage for an SiPM of type Hamamatsu S13360-6025PE~\cite{HamamatsuS13360}. \emph{Top:} The measured signal. \emph{Bottom:} The signal sizes divided by the overvoltage. All curves are normalized to a PDE of 1 at $\Vov=\SI{5}{\volt}$. Taken from~\cite{ThesisJulian}.}
    \label{fig:PDEMeasurement}
\end{figure}

The setup presented in section~\ref{sec:DynamicRangeMeasurement} was used for this measurement with the identical SiPM of type Hamamatsu S13360-6025PE as used in all other measurements. The brightness of the pulsed light source was set so that $\lesssim\SI{10}{\percent}$ of the SiPM cells were triggered. The rather low signal reduced the amount of cells that were hit by at least two photons to about \SI{20}{\percent}. The raw signal is shown in the top plot of figure~\ref{fig:PDEMeasurement}. To correct for a systematic offset, the average signal of the first four data points is shifted to zero.

The measured signal reaches zero at an overvoltage below zero. The breakdown voltage of individual cells differs across the SiPM resulting in a smeared region around the average breakdown voltage. The variation of breakdown voltages is assumed to be Gaussian and is deconvoluted from the measurement using the algorithm presented in section~\ref{sec:IdealSiPMIncidentPhotonDistribution}. The width $\sigma$ of the Gaussian is unknown and was varied between $\SI{0.05}{\volt}\leq\sigma\leq\SI{0.2}{\volt}$ which is a reasonable range. A width of $\sigma=\SI{0.17}{\volt}$ corresponding to \SI{3.4}{\percent} of the recommended overvoltage of \SI{5}{\volt} was found to yield good results.

In the bottom plot of figure~\ref{fig:PDEMeasurement}, the relative PDE is shown when applying equation~\eqref{eq:PDESignalOV} and normalizing the PDE to 1 at $\Vov=\SI{5}{\volt}$.
The difference between the raw data and the deconvoluted one is small, meaning that the exact width of the Gaussian has only a low impact on the result.
Different functions such as a single exponential which was suggested in~\cite{Otte:2016aaw} were fit to the result but do not describe the data over the full range of overvoltages. While they succeed in the region of $\Vov>\SI{1}{\volt}$ they fail at lower overvoltage. Only a function of the following form was found to describe the data over the full range
\begin{equation}\label{eq:PDERecoveryFit}
	\textit{PDE}\left(V_\text{ov}\right) = c_0\cdot\left(\tanh\left(\frac{V_\text{ov}-c_1}{c_2}\right)+1\right)\cdot{}\left(1-c_5\cdot e^{-\frac{V_\text{ov}-c_1}{c_3}}-\left(1-c_5\right) e^{-\frac{V_\text{ov}-c_1}{c_4}}\right)\\
\end{equation}
with the values for the parameters $c_i$ given in table~\ref{tab:PDERecoveryParameters}.
\begin{table}\centering
\begin{tabular}{|c|c|c|c|c|c|c|}\hline
	parameter & $c_0$ & $c_1$ / V & $c_2$ / V & $c_3$ / V & $c_4$ / V & $c_5$\\\hline
	value & 1.27 & \num{5.9e-3} & 1.62 & 15.3 & 0.191 & 0.838\\\hline
	$\sigma_{c_i}$ & 0.10 & \num{32.7e-3} & 0.12 & 1.9 & 0.035 & 0.010\\\hline
\end{tabular}
	\caption{Resulting parameters and their standard deviation for the fit function given in equation \eqref{eq:PDERecoveryFit}. The corresponding function is shown in figure~\ref{fig:PDEMeasurement}.}
	\label{tab:PDERecoveryParameters}
\end{table}%
\newpage

\section{Pseudo code of the SiPM simulation}\label{app:PseudoCode}
All variable names follow the convention introduced in section~\ref{sec:SiPMElectricModel}. The variable $N_\mathrm{edge}$ refers to the number of cells along one side of the SiPM.

\begin{algorithm}[H]
    \SetAlgoLined
    \DontPrintSemicolon
    \SetKwProg{Cl}{Class}{:}{}
    \SetKwProg{Fn}{Function}{:}{}
    \SetKwProg{Const}{Constructor}{:}{}
    \SetKwFunction{SiPM}{SiPM}%
    \SetKwFunction{GetVone}{GetV1}%
    \SetKwFunction{GetVtwo}{GetV2}%
    \SetKwFunction{GetVthree}{GetV3}%
    \SetKwFunction{UpdateVone}{UpdateV1}%
    \SetKwFunction{HitCell}{HitCell}%
    \Cl{\SiPM{}}{
        \Const{\SiPM{$N_\mathrm{edge}$, $\Rq$, $\Rs$, $\Cd$, $\Cq$, $\Cg$, $\Vov$, $\pxt$}}{
            $Q \leftarrow$ eq.~\eqref{eq:SiPMGain}\;
            $\tau_\pm, \tau_\mathrm{q}, A_1, A_2, c_1, c_2 \leftarrow$ eq.~\eqref{eq:SiPMTimeConstants} \& \eqref{eq:SiPMRecoveryConstants}\;\;
            
            Declare quadratic array with number of cells to store the time of the last hit for each cell individually.\;
            $\textit{lastHit}[N_\mathrm{edge}][N_\mathrm{edge}] \leftarrow 0$\;\;
            
            Time of the last photon that hit the SiPM.\;
            $t_\mathrm{last} \leftarrow 0$\;\;
            
            Initialize time evolution $\tilde{\Vone}$, $\tilde{\Vtwo}$, $\Vthree$ as zero.\;
            The time evolution of $\tilde{\Vone}$ needs to be determined for each cell individually.\;
            $\tilde{\Vone}[N_\mathrm{edge}][N_\mathrm{edge}], \tilde{\Vtwo}, \Vthree \leftarrow 0$\;\;
            
            Initialize factors in front of the exponential functions in equation~\eqref{eq:SiPMTimeConstants}.\;
            Indices $1,2,3$ refer to the voltages $\Vone$, $\Vtwo$, $\Vthree$.\;
            Indices $q,+,-$ refer to the factors in front of the corresponding exponential function.\;
            The factors for $\Vone$ need to be calculated for each cell individually.\;
            $A_{1,q}[N_\mathrm{edge}][N_\mathrm{edge}], A_{1,+}[N_\mathrm{edge}][N_\mathrm{edge}], A_{1,-}[N_\mathrm{edge}][N_\mathrm{edge}] \leftarrow 0$\;
            $A_{2,q}, A_{2,+}, A_{2,-} \leftarrow 0$\;
            $A_{3,+}, A_{3,-} \leftarrow 0$
        }\;
        
        \Fn{\GetVone{$N_\mathrm{x}$, $N_\mathrm{y}$, $\Delta{}t$}}{
            $\tilde{\Vone} \leftarrow A_{1,q}[N_\mathrm{x}][N_\mathrm{y}]\cdot\exp(-\Delta{}t/\tau_\mathrm{q})$\;
            $\tilde{\Vone} \leftarrow \tilde{\Vone}+A_{1,+}[N_\mathrm{x}][N_\mathrm{y}]\cdot\exp(-\Delta{}t/\tau_\mathrm{+})$\;
            $\tilde{\Vone} \leftarrow \tilde{\Vone}+A_{1,-}[N_\mathrm{x}][N_\mathrm{y}]\cdot\exp(-\Delta{}t/\tau_\mathrm{-})$\;
            \KwRet $\tilde{\Vone}$
        }\;
        
        \Fn{\UpdateVone{$N_\mathrm{x}$, $N_\mathrm{y}$, $\Delta{}t$, $Q$}}{
            $A_{1,q}[N_\mathrm{x}][N_\mathrm{y}] \leftarrow A_{1,q}[N_\mathrm{x}][N_\mathrm{y}]\cdot\exp(-\Delta{}t/\tau_\mathrm{q})-\frac{Q\Rq}{N}\cdot\frac{N-1}{\tau_\mathrm{q}}$ (eq.~\eqref{eq:SiPMVoltageRecovery})\;
            $A_{1,+}[N_\mathrm{x}][N_\mathrm{y}] \leftarrow A_{1,+}[N_\mathrm{x}][N_\mathrm{y}]\cdot\exp(-\Delta{}t/\tau_+)-\frac{Q\Rq}{N}\frac{\Rs\Cg}{c_2}\cdot\left(1-A_1\right)$ (eq.~\eqref{eq:SiPMVoltageRecovery})\;
            $A_{1,-}[N_\mathrm{x}][N_\mathrm{y}] \leftarrow A_{1,-}[N_\mathrm{x}][N_\mathrm{y}]\cdot\exp(-\Delta{}t/\tau_-)-\frac{Q\Rq}{N}\frac{\Rs\Cg}{c_2}\cdot A_1$ (eq.~\eqref{eq:SiPMVoltageRecovery})\;
        }
    }
\end{algorithm}
\newpage
\begin{algorithm}[H]
    \SetAlgoLined
    \DontPrintSemicolon
    \SetKwProg{Cl}{Class}{:}{}
    \SetKwProg{Fn}{Function}{:}{}
    \SetKwFunction{SiPM}{SiPM}%
    \SetKwFunction{GetVone}{GetV1}%
    \SetKwFunction{GetVtwo}{GetV2}%
    \SetKwFunction{GetVthree}{GetV3}%
    \SetKwFunction{UpdateVone}{UpdateV1}%
    \SetKwFunction{UpdateVtwo}{UpdateV2}%
    \SetKwFunction{UpdateVthree}{UpdateV3}%
    \SetKwFunction{HitCell}{HitCell}%
    \SetKwFunction{Random}{Random}%
    \SetKwFunction{PDE}{PDE}%
    \SetKwBlock{Begin}{}{}
    \Begin{
        \Fn{\GetVtwo{$t$}}{
            $\Delta{}t \leftarrow t-t_\mathrm{last}$\;
            $\tilde{\Vtwo} \leftarrow A_{2,q}\cdot\exp(-\Delta{}t/\tau_\mathrm{q})$\;
            $\tilde{\Vtwo} \leftarrow \tilde{\Vtwo}+ A_{2,+}\cdot\exp(-\Delta{}t/\tau_\mathrm{+})$\;
            $\tilde{\Vtwo} \leftarrow \tilde{\Vtwo}+ A_{2,-}\cdot\exp(-\Delta{}t/\tau_\mathrm{-})$\;
            \KwRet $\tilde{\Vtwo}$
        }\;
        
        \Fn{\UpdateVtwo{$t$, $Q$}}{
            $\Delta{}t \leftarrow t-t_\mathrm{last}$\;
            $A_{2,q} \leftarrow A_{2,q}\cdot\exp(-\Delta{}t/\tau_\mathrm{q})+\frac{Q\Rq}{N\tau_\mathrm{q}}$ (eq.~\eqref{eq:SiPMVoltageRecovery})\;
            $A_{2,+} \leftarrow A_{2,+}\cdot\exp(-\Delta{}t/\tau_+)-\frac{Q\Rq}{N}\frac{\Rs\Cg}{c_2}\cdot\left(1-A_1\right)$ (eq.~\eqref{eq:SiPMVoltageRecovery})\;
            $A_{2,-} \leftarrow A_{2,-}\cdot\exp(-\Delta{}t/\tau_-)-\frac{Q\Rq}{N}\frac{\Rs\Cg}{c_2}\cdot A_1$ (eq.~\eqref{eq:SiPMVoltageRecovery})\;
        }\;
        
        \Fn{\GetVthree{$t$}}{
            $\Delta{}t \leftarrow t-t_\mathrm{last}$\;
            $\Vthree \leftarrow A_{3,+}\cdot\exp(-\Delta{}t/\tau_\mathrm{+})$\;
            $\Vthree \leftarrow \Vthree+ A_{3,-}\cdot\exp(-\Delta{}t/\tau_\mathrm{-})$\;
            \KwRet $\Vthree$
        }\;
        
        \Fn{\UpdateVthree{$t$, $Q$}}{
            $\Delta{}t \leftarrow t-t_\mathrm{last}$\;
            $A_{3,+} \leftarrow A_{3,+}\cdot\exp(-\Delta{}t/\tau_+)+\frac{Q\Rs\Rq\Cq}{c_2}\cdot\left(1-A_2\right)$ (eq.~\eqref{eq:SiPMVoltageRecovery})\;
            $A_{3,-} \leftarrow A_{3,-}\cdot\exp(-\Delta{}t/\tau_-)+\frac{Q\Rs\Rq\Cq}{c_2}\cdot A_2$ (eq.~\eqref{eq:SiPMVoltageRecovery})\;
        }
    }
\end{algorithm}
\newpage
\begin{algorithm}[H]
    \SetAlgoLined
    \DontPrintSemicolon
    \SetKwProg{Cl}{Class}{:}{}
    \SetKwProg{Fn}{Function}{:}{}
    \SetKwFunction{SiPM}{SiPM}%
    \SetKwFunction{GetVone}{GetV1}%
    \SetKwFunction{GetVtwo}{GetV2}%
    \SetKwFunction{GetVthree}{GetV3}%
    \SetKwFunction{UpdateVone}{UpdateV1}%
    \SetKwFunction{UpdateVtwo}{UpdateV2}%
    \SetKwFunction{UpdateVthree}{UpdateV3}%
    \SetKwFunction{HitCell}{HitCell}%
    \SetKwFunction{Random}{Random}%
    \SetKwFunction{PDE}{PDE}%
    \SetKwBlock{Begin}{}{}
    \Begin{
        Function to calculate the effect when a cell is hit.\;
        $N_\mathrm{x}$ and $N_\mathrm{y}$ declare the index of the hit cell, $t$ is the time of the hit.\;
        \Fn{\HitCell{$N_\mathrm{x}$, $N_\mathrm{y}$, $t$}}{
            $\Delta{}t \leftarrow t-\textit{lastHit}[N_\mathrm{x}][N_\mathrm{y}]$\;\;
            
            Calculate the instantaneous overvoltage\;
            $\Vovinst \leftarrow \Vbi+\GetVone{$\Delta{}t$}+\GetVtwo{t}-\GetVthree{t}$\;\;
            
            Random decision according to PDE (eq.~\eqref{eq:PDERecoveryFit}) if avalanche is triggered.\;
            \If{\Random{0,1}>\PDE{$\Vovinst$}}{
                \KwRet 0\;
            }\;
            
            Calculate released charge $\tilde{Q}$.\;
            $\tilde{Q} \leftarrow \Vovinst/\Vov\cdot{}Q$\;\;
            
            Update the amplitudes of the exponential functions.\;
            \UpdateVone{$N_\mathrm{x}$, $N_\mathrm{y}$, $\Delta{}t$, $\tilde{Q}$}\;
            \UpdateVtwo{$t$, $\tilde{Q}$}\;
            \UpdateVthree{$t$, $\tilde{Q}$}\;\;
            
            Update time of the last photon hit.\;
            $\textit{lastHit}[N_\mathrm{x}][N_\mathrm{y}] \leftarrow t$\;
            $t_\mathrm{last} \leftarrow t$\;\;
            
            Calculate instantaneous probability to emit a photon that can produce crosstalk.\;
            $\tilde{p}_\mathrm{xt} \leftarrow \Vovinst/\Vov\cdot\pxt$\;
            \While{\Random{0,1}<$\tilde{p}_\mathrm{xt}$}{
                $\textit{cell} \leftarrow \Random{0,3}$\;
                \uCase{$\textit{cell}==0$}{$\tilde{Q} \leftarrow \tilde{Q} +\HitCell{$N_\mathrm{x}-1$,$N_\mathrm{y}$,$t$}$}
                \uCase{$\textit{cell}==1$}{$\tilde{Q} \leftarrow \tilde{Q} +\HitCell{$N_\mathrm{x}+1$,$N_\mathrm{y}$,$t$}$}
                \uCase{$\textit{cell}==2$}{$\tilde{Q} \leftarrow \tilde{Q} +\HitCell{$N_\mathrm{x}$,$N_\mathrm{y}-1$,$t$}$}
                \uCase{$\textit{cell}==3$}{$\tilde{Q} \leftarrow \tilde{Q} +\HitCell{$N_\mathrm{x}$,$N_\mathrm{y}+1$,$t$}$}
            }\;
            \KwRet $\tilde{Q}$
        }
    }
\end{algorithm}

\section{List of symbols}\label{app:ListOfSymbols}
\begin{tabular}{p{.1\textwidth}p{.9\textwidth}}
    $f_i$ & The charge in units of p.e.~released at the SiPM when hit by a photon.\\
    $\SOft$ & Time dependent voltage signal produced by the SiPM.\\
    $\Vspe$ & Amplitude of the time dependent voltage produced in the breakdown of a single fully recovered cell.\\
    $\SPEOft$ & Shape of the time dependent voltage signal produced in the breakdown of a single fully recovered cell.\\
    $\SPERealOft$ & Shape of the time dependent voltage signal produced in the breakdown of a single cell. It has the same shape as $\SPEOft$ but is scaled by a factor $f_i$.\\
    $\SPEOftAndTau$ & Probability distribution function of measuring a voltage signal at time $t$ if a photon hit the SiPM at time $\tau$.\\
    $\textit{SPE}'(\tau|t)$ & Probability distribution function that a photon hit the sensor at time $\tau$ if a voltage signal was measured at time $t$.\\
    $\Nmeas$ & Total measured signal at the SiPM in terms of p.e.\\
    $N_\mathrm{inc}$ &  Total number of photons incident on the SiPM.\\
    $\NgammaOft$ & The incident equivalent distribution of photons impinging on a real SiPM. The term \emph{equivalent} refers to the fact that a PDE of \SI{100}{\percent} is assumed.\\
    $\NgammaPrimeOft$ & The incident distribution of photons for an ideal SiPM. The term \emph{ideal} refers to the fact that all cells are always fully recovered so that each photon initiates an avalanche.\\
    $N_{\gamma,0}(t)$ & The incident equivalent distribution of photons impinging on the real SiPM at the start of the iteration process. It equals $\NgammaPrimeOft$.\\
    $N_{\gamma,k}(t)$ & The incident equivalent distribution of photons impinging on the real SiPM after $k$ iterations.\\
    $N_\gamma^w(t)$ & The time distribution when each photon in $N_\gamma(t)$ is weighted with the corresponding released charge $f_i$ of the SiPM.\\
    $N_{\gamma,0}^w(t)$ & The time distribution when each photon in $\NgammaPrimeOft$ is weighted with the corresponding released charge $f_i$ of the SiPM.\\
    $N_{\gamma,k}^w(t)$ & The time distribution when each photon in $N_{\gamma,k}(t)$ is weighted with the corresponding released charge $f_i$ of the SiPM.\\
    $N_{\gamma,\mathrm{sim}}(t)$ & The simulated incident equivalent photon time distribution.\\
    $R_k(t)$ & The ratio of the time distribution of photons impinging on the ideal SiPM $\NgammaPrimeOft$ with the weighted distribution $N_{\gamma,k}^w(t)$ after $k$ iterations.\\
    $\noiseOft$ & Time dependent random noise on the measured voltage signal.\\
    $M$ & Response matrix of the ideal SiPM. Converts the time dependent incident photon distribution $\NgammaPrimeOft$ to the measured voltage signal $\SOft$.\\
    $P(t)$ & Probability for a photon to hit the SiPM at time $t$.\\
\end{tabular}

\end{document}

%% file: img/SiPMCircuitDiagram.tex
\begin{circuitikz}[scale=0.9]
	\node[anchor=west] at (0,0) {$V_\mathrm{bi}$};
	\draw (0,0) to[short,f=$I_\mathrm{bi}$,o-] (0,-1.5) node[circ]{} to[short] (-6,-1.5)
	      to[short] (-6,-3) node[circ]{}
	      to[short] (-7,-3)
	      to[R,l_=$R_\mathrm{q}$] (-7,-5)
	      to[short] (-6,-5) node[circ]{}
	      (-6,-3) to[short] (-5,-3)
	      to[C,l=$C_\mathrm{q}$] (-5,-5)
	      to[short] (-6,-5)
	      to[short,f=$I_\mathrm{q}$,l_=$\Vone$] (-6,-7) node[circ]{}
	      to[short] (-7,-7)
	      to[american current source,l_=$I_\mathrm{d}$] (-7,-9)
	      to[short] (-6,-9) node[circ]{}
	      (-6,-7) to[short,f_<=$I_\mathrm{p}$] (-4.5,-7)
	      to[C,l=$C_\mathrm{d}$] (-4.5,-9)
	      to[short] (-6,-9)
	      to[short] (-6,-10)
	      to[short] (0,-10) node[circ]{}

		  (0,-1.5) to[short,f<=$I_\mathrm{t}$] (0,-3) node[circ]{}
		  to[short] (-1,-3) 
		  to[R,l_=$\frac{R_\mathrm{q}}{N-1}$] (-1,-5)
		  to[short] (0,-5) node[circ]{}
		  (0,-3) to[short] (1,-3)
		  to[C,l=$\left(N-1\right)C_\mathrm{q}$] (1,-5)
		  to[short] (0,-5)
		  to[short] (0,-5)
	      to[short,f<=$I_\mathrm{u}$,l_=$\Vtwo$] (0,-7)
	      to[C,l=$\left(N-1\right)C_\mathrm{d}$] (0,-9)
	      to[short] (0,-10)
	      
		  (0,-1.5) to[short] (6,-1.5)
		  to[short,f<=$I_\mathrm{g}$] (6,-5)
		  to[C,l=$C_\mathrm{g}$] (6,-7)
		  to[short] (6,-10)
		  to[short] (0,-10)
		  
		  to[short,f=$I_\mathrm{SiPM}$,l_=$\Vthree$] (0,-11.5) node[circ]{}
		  to[R,l=$R_\mathrm{s}$] (-2,-11.5)
		  node[ground,rotate=-90] {}
		  (0,-11.5) to[short,-o] (0,-12.5) node[anchor=west] {$V_\mathrm{meas}$};
		  
	\draw[line width=1,dashed] (-8.5,-10.2) rectangle (-3.1,-1.2);
	\node[anchor=south] at (-5.8,-1.2) {Triggered cell};
	\draw[line width=1,dashed] (-2.9,-10.2) rectangle (4,-1.2);
	\node[anchor=south west] at (1,-1.2) {Untriggered cells};
\end{circuitikz}

%% file: img/SiPMSimulationWorkflow.tex
\begin{tikzpicture}[every node/.style={draw=black,text=black,text width=0.25\textwidth},align=center]

\node (start) [rectangle] {Photon impinges on cell at time $\tau$};
\node (voltageCheck) [rectangle,below left=0.05\textwidth and 0.1\textwidth of start] {Determine $\Vone'=\sum_{t_i<\tau}\tilde{\Vone}(t_i)$};
\node (voltageCheck2) [rectangle,below=0.05\textwidth of start] {Determine $\Vtwo'=\sum_{t_j<\tau}\tilde{\Vtwo}(t_j)$};
\node (readoutVoltage) [rectangle,below right=0.05\textwidth and 0.1\textwidth of start] {Determine $\Vthree'=\sum_{t_{i,j}<\tau}\tilde{\Vthree}(t_{i,j})$};
\node (overVoltage) [rectangle,below=0.05\textwidth of voltageCheck2,text width=0.4\textwidth] {Determine applied overvoltage $\tilde{V}_\mathrm{ov}=\Vbi+\Vone'+\Vtwo'-\Vthree'$};

\node (PDE) [rectangle,below=0.05\textwidth of overVoltage] {Check PDE and if avalanche is triggered};

\node (NoDetection) [rectangle,right=0.1\textwidth of PDE] {Nothing happens};

\node (Detection) [rectangle,below=0.05\textwidth of PDE] {Determine\\ released charge $\tilde{Q}\propto\tilde{V}_\mathrm{ov}$ (eq.~\eqref{eq:SiPMGain})};


\node (Crosstalk) [rectangle,left=0.1\textwidth of Detection] {Check if crosstalk occurs};

\path let \p1 = (start), \p2 = (voltageCheck) in coordinate (Corner) at  (\x2,\y1);

\draw[->,black,shorten <= 3pt,shorten >= 3pt] (start) -- node[above left,fill=none,draw=none,text width=0.04\textwidth] {eq.~\eqref{eq:SiPMVoltageRecovery}} (voltageCheck);
\draw[->,black,shorten <= 3pt,shorten >= 3pt] (start) -- node[right,fill=none,draw=none,text width=0.04\textwidth] {eq.~\eqref{eq:SiPMVoltageRecovery}} (voltageCheck2);
\draw[->,black,shorten <= 3pt,shorten >= 3pt] (start) -- node[above right,fill=none,draw=none,text width=0.04\textwidth] {eq.~\eqref{eq:SiPMVoltageRecovery}} (readoutVoltage);

\draw[->,black,shorten <= 3pt,shorten >= 3pt] (voltageCheck) -- (overVoltage);
\draw[->,black,shorten <= 3pt,shorten >= 3pt] (voltageCheck2) -- (overVoltage);
\draw[->,black,shorten <= 3pt,shorten >= 3pt] (readoutVoltage) -- (overVoltage);

\draw[->,black,shorten <= 3pt,shorten >= 3pt] (overVoltage) -- (PDE);

\draw[->,black,shorten <= 3pt,shorten >= 3pt] (PDE) -- node[above,fill=none,draw=none] {no} (NoDetection);

\draw[->,black,shorten <= 3pt,shorten >= 3pt] (PDE) -- node[left,fill=none,draw=none,text width=0.04\textwidth] {yes} (Detection);

\draw[->,black,shorten <= 3pt,shorten >= 3pt] (Detection) -- (Crosstalk);

\draw[black,shorten <= 3pt,shorten >= 3pt] (Crosstalk) -- node[left,fill=none,draw=none,text width=0.04\textwidth] {yes} (voltageCheck);
\draw[black,shorten <= 3pt] (voltageCheck) -- (Corner);
\draw[->,black,shorten >= 3pt] (Corner) -- (start);
\end{tikzpicture}

%% file: img/SiPMDeconvolutionSchematic.tex
\begin{tikzpicture}[every node/.style={draw=black,text width=0.25\textwidth},align=center]
	
	\node (signal) [rectangle,color=MPLOrange,text=black,line width=1.5] {measured signal $S\left(t\right)$};
	\node [above=0.05\textwidth of signal,draw=none] (Inlet1) {\includegraphics[width=\textwidth]{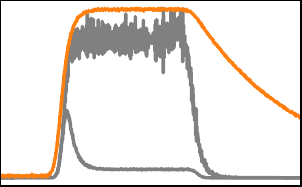}};
	\draw[<->,shorten <= 3pt,shorten >= 3pt] (signal) -- (Inlet1);
	
	\node (ideal) [rectangle,right=0.35\textwidth of start,color=MPLBlue,text=black,line width=1.5] {incident photon time distribution $N'_\gamma\left(t\right)$};
	\draw[->,shorten <= 3pt,shorten >= 3pt] (signal) -- (ideal) node (deconvolution) [draw=none,fill=none,midway,anchor=south] {deconvolution of $S\left(t\right)$};
	
	\node [right=0.35\textwidth of Inlet1,draw=none] (Inlet2) {\includegraphics[width=\textwidth]{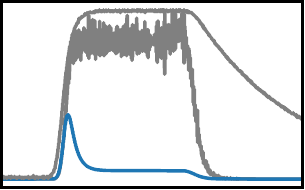}};
	\draw[<->,shorten <= 3pt,shorten >= 3pt] (ideal) -- (Inlet2);
	
	\node (assumptions) [rectangle,above=0.05\textwidth of deconvolution] {Assume no saturation effects, $\textit{PDE}=1$};
	\draw[->,shorten <= 3pt,shorten >= 3pt] (assumptions) -- (deconvolution);
	
	\node (estimate) [rectangle,below=0.07\textwidth of ideal] {estimate for incident equivalent photon distribution $N_{\gamma,k}\left(t\right)$};
	\draw[->,shorten <= 3pt,shorten >= 3pt] (ideal) -- (estimate);
	
	\node (weighted) [rectangle,left=0.35\textwidth of estimate] {$N_{\gamma,k}\left(t\right)$ weighted with the SiPM response $\Rightarrow$ $N^w_{\gamma,k}\left(t\right)$};
	\draw[->,shorten <= 3pt,shorten >= 3pt] (estimate) -- (weighted) node (sipmsimulation) [draw=none,fill=none,midway,anchor=south] {simulate SiPM response for $N_{\gamma,k}\left(t\right)$};
	
	\node (check) [rectangle,below=0.1\textwidth of weighted] {Check if $N^w_{\gamma,k}\left(t\right)=N'_\gamma\left(t\right)$};
	\draw[->,shorten <= 3pt,shorten >= 3pt] (weighted) -- (check);
	
	\node (checkYes) [rectangle,below=0.07\textwidth of check,color=MPLGreen,text=black,line width=1.5] {Incident equivalent photon distribution $N_\gamma=N^w_{\gamma,k}$};
	\draw[->,shorten <= 3pt,shorten >= 3pt] (check) -- (checkYes) node (yes) [fill=none,draw=none,anchor=west,midway,text width=width("yes")] {yes};
	
	\node [right=0.05\textwidth of checkYes,draw=none] (Inlet3) {\includegraphics[width=\textwidth]{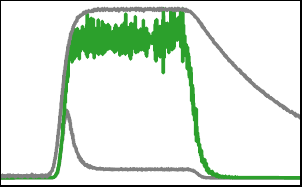}};
	\draw[<->,shorten <= 3pt,shorten >= 3pt] (checkYes) -- (Inlet3);
	
	\node (checkNo) [rectangle,right=0.35\textwidth of check] {$R_k\left(t\right)=\frac{N'_\gamma\left(t\right)}{N^w_{\gamma,k}\left(t\right)}$};
	\draw[->,shorten <= 3pt,shorten >= 3pt] (check) -- (checkNo) node (no) [fill=none,draw=none,anchor=south,midway] {no};
	
	\draw[->,shorten <= 3pt,shorten >= 3pt] (checkNo) -- (estimate) node (update) [fill=none,draw=none,anchor=east,midway,text width=width("{$R_k\left(t\right)\cdot N_{\gamma,k}\left(t\right)$}")] {$N_{\gamma,k+1}\left(t\right)=R_k\left(t\right)\cdot N_{\gamma,k}\left(t\right)$};
\end{tikzpicture}